\documentclass{article}

\usepackage{preprint}

\usepackage{amsmath, amsthm, amssymb, amsfonts}

\usepackage[utf8]{inputenc}	
\usepackage[T1]{fontenc}	  
\usepackage{xcolor}		      
\usepackage{booktabs} 	
\usepackage{mathptmx}
\usepackage{nicefrac}		    
\usepackage{microtype}		  
\usepackage{afterpage}
\usepackage{graphicx}
\usepackage{float}			    
\usepackage{array}          
\usepackage{booktabs}
\usepackage{placeins}

\usepackage{citation-style-language}
\cslsetup{style = asa}
\usepackage{caption}

\DeclareCaptionFormat{boldnumber}{\textbf{#1#2}#3}
\usepackage{soul}

\usepackage{titlesec}
\titlespacing\section{0pt}{12pt plus 3pt minus 3pt}{1pt plus 1pt minus 1pt}
\titlespacing\subsection{0pt}{10pt plus 3pt minus 3pt}{1pt plus 1pt minus 1pt}
\titlespacing\subsubsection{0pt}{8pt plus 3pt minus 3pt}{1pt plus 1pt minus 1pt}

\titleformat{\section}
  {\normalfont\Large}{\thesection}{1em}{\MakeUppercase}
\titleformat{\subsection}
  {\normalfont\large\bfseries}{\thesubsection}{1em}{}
\titleformat{\subsubsection}
  {\normalfont\normalsize\itshape}{\thesubsubsection}{1em}{}

{\sffamily \footnotesize \begin{tabular}{#1}}%
{\end{tabular}}

\let\oldtable\table
\let\endoldtable\endtable
\renewenvironment{table}[1][!htbp]{ 
  \oldtable[#1]
  \footnotesize 
}{
  \endoldtable
}

\usepackage[hyperindex=true, colorlinks=true, linkcolor=blue, urlcolor=blue, citecolor=blue, pdfborder={0 0 0}]{hyperref}

\usepackage{tikz}

\definecolor{lime}{HTML}{A6CE39}
\DeclareRobustCommand{\orcidicon}{
	\begin{tikzpicture}
	\draw[lime, fill=lime] (0,0)
	circle [radius=0.16]
	node[white] {{\fontfamily{qag}\selectfont \tiny ID}};
	\draw[white, fill=white] (-0.0625,0.095)
	circle [radius=0.007];
	\end{tikzpicture}
	\hspace{-2mm}
}
\foreach \x in {A, ..., Z}{\expandafter\xdef\csname orcid\x\endcsname{\noexpand\href{https://orcid.org/\csname orcidauthor\x\endcsname}
			{\noexpand\orcidicon}}
}

\title{In Silico Sociology: Forecasting COVID-19 Polarization with Large Language Models}

\usepackage{authblk}

\author[1]{Austin C. Kozlowski\orcidA{}}
\author[2]{Hyunku Kwon\orcidB{}}
\author[3]{James A. Evans\orcidC{}}

\affil[1, 2, 3]{Department of Sociology, University of Chicago}
\affil[3]{Santa Fe Institute}

\begin{document}

  \maketitle

  \begin{abstract}
    By training deep neural networks on massive archives of digitized text,
    large language models (LLMs) learn the complex linguistic patterns that
    constitute historic and contemporary discourses. We argue that LLMs can
    serve as a valuable tool for sociological inquiry by enabling accurate
    simulation of respondents from specific social and cultural contexts.
    Applying LLMs in this capacity, we reconstruct the public opinion
    landscape of 2019 to examine the extent to which the future polarization
    over COVID-19 was prefigured in existing political discourse. Using an
    LLM trained on texts published through 2019, we simulate the responses
    of American liberals and conservatives to a battery of pandemic-related
    questions. We find that the simulated respondents reproduce observed
    partisan differences in COVID-19 attitudes in 84\% of cases,
    significantly greater than chance. Prompting the simulated respondents
    to justify their responses, we find that much of the observed partisan
    gap corresponds to differing appeals to freedom, safety, and
    institutional trust. Our findings suggest that the politicization of
    COVID-19 was largely consistent with the prior ideological landscape,
    and this unprecedented event served to advance history along its track
    rather than change the rails.
  \end{abstract}
  \vspace{0.35cm}

\section*{INTRODUCTION}

For decades, the term ``artificial intelligence'' was used to describe
computational capabilities that remained out of reach---in a quote often
attributed to Larry Tesler, ``artificial intelligence is whatever hasn't
been done yet.''\footnote{Tesler himself maintains that this is a
  misquote, and that his actual statement was ``intelligence is whatever
  machines haven't done yet,'' critiquing the tendency
  to continuously shift the criteria for intelligence to keep it one
  step ahead of ever-encroaching machines \parencite{Tesler2010-ty}.} But in
recent years, algorithms have gained fluency in such complex tasks as
composing novel texts, generating photo-realistic images, and advanced
coding, and the term ``artificial intelligence'' is now part of daily
parlance.

This rapid progress is largely the product of two overarching
technological developments. First, continuous improvements to hardware
have driven decades of exponential growth in computational power. New
capabilities emerge as models scale up, and modern chips now make it
possible to train models with over a trillion parameters \parencite{Kaplan2020-lo}.
Second, the proliferation of online digital content has supplied
abundant training data. Current models are commonly trained on a
near-complete record of all text on the internet, and ``multi-modal''
models are trained on vast collections of online images and videos as
well \parencite{Brown2020-nh}. These two complementary developments have
resulted in algorithms capable of generating vast and varied forms of
content; deep neural architectures make it possible to learn subtle and
complex patterns, and large training data provide abundant examples of
patterns to learn.

Particularly striking advances have been made in the training of large
language models (LLMs), algorithms capable of generating text by
predicting the next word in a sequence. LLMs form the foundation of AI
conversational agents such as OpenAI's ChatGPT, Anthropic's Claude, and
Deep Mind's Gemini. These ``chatbots'' have swiftly gained widespread
public exposure; ChatGPT alone reached over 100 million users within two
months of its public release, and a large share of workers in fields
ranging from education to computer programming report regularly using
LLMs to improve their productivity
\parencite{DellAcqua2023-fo, Mollick2023-wy}. Seemingly overnight, the ability of
algorithms to successfully impersonate human interaction -- commonly
known as ``the Turing Test'' \parencite{Turing1950-in} -- shifted
from aspiration to expectation.

Because LLMs are typically trained on the wide variety of texts
published on the internet, they learn to reproduce many distinct
discursive styles. They achieve this not by memorizing specific
sentences (although they occasionally do this), but by learning the
latent probability distributions of word sequences constitutive of
discourses. Publicly released LLMs are commonly fine-tuned to speak in
the style of a helpful professional assistant and avoid making
statements that are offensive, biased, or politically charged
\parencite{Ouyang2022-xw}. But if
prompted to do so, even these fine-tuned models can generate texts that
mirror the diverse cultural and linguistic styles represented in their
training data, ranging from sarcastic wisecracking to postmodern
literary criticism to extreme political rhetoric \parencite{Argyle2023-ii, Kim2023-ji, Park2022-py}.

We argue that this capacity to mimic and reproduce human responses opens
fruitful new forms of socio-cultural analysis. First, LLMs are able to
reproduce the discourses of populations not available for interviews or
surveys, including populations from the past. Second, because LLMs can
quickly and cheaply generate responses, they facilitate testing wide
varieties of wordings for each question, improving robustness and
identifying how specific words and associated framings can steer
responses \parencite{Garcia-Pardina2022-cx}. Third, they allow open-ended responses to be both generated
and effectively machine-coded at scale, enabling a high-level view into
the system of considerations that inform a given response. Establishing
the validity of these LLM methods enables the generation of useful
``social science fictions'' or simulations that open up a wide range of
possibilities that lie beyond the scope of this empirical investigation.

We apply this novel approach to investigate a longstanding question in
the study of culture -- to what extent are new cultural developments
constrained by the existing ideological landscape
\parencite{Converse1964-kv, DellaPosta2020-ta, Hunzaker2019-ds}? More specifically, when
a new issue arises, are public responses to that issue predictable given
the systematization of attitudes across other topics? Answering this
question requires observing how individuals respond to an emerging issue
\emph{before it is framed} by commentators, public figures, or personal
acquaintances. Yet this analytic approach presents an empirical
challenge, as new issues are rapidly subject to public discussion and
interpretation, leaving social scientists little opportunity to measure
responses prior to top-down framing by opinion leaders.

LLMs can shed new light on this question by serving as a ``cultural time
capsule,'' capable of reproducing the most plausible responses from the
time of the model's training texts. An analyst can present an LLM with
issues that \emph{had not yet emerged when the model was trained} and
compare these LLM-generated responses with survey responses gathered in
the following years. To the extent that the LLM generates the pattern of
attitudes that empirically manifest in later years, it suggests that
public response to the new issue was prefigured in prior discourse, and
that the public reception of the issue was a ``predictable''
development.

We use this method to examine a pivotal issue of our time, the public
response to the COVID-19 pandemic. The spread of COVID-19 foregrounded
several topics with little political precedent, such as vaccine
mandates, face masks, and lockdowns. The public response to these issues
rapidly politicized, with liberals endorsing cautious approaches to the
virus and conservatives opposing more drastic measures
\parencite{Gadarian2021-su}. What remains unclear is whether this politicization
reflects deep characteristics of American liberalism and conservatism as
idea systems, or whether the public simply responded to the cues of
partisan elites who quickly espoused opposing responses. Fortunately for
our analytic aims, GPT-3, the first LLM to accurately reproduce patterns
of public opinion, was trained on texts published through October 2019,
and therefore has no knowledge of the COVID-19 pandemic
\parencite{OpenAI2023-wo}. We can
therefore use GPT-3 to reproduce how liberals and conservatives
\emph{likely would have responded} to questions pertaining to a pandemic
before any top-down framing emerged around COVID-19
\parencite{Kaplan2008-nm}.

We find that liberal and conservative responses simulated with GPT-3
largely anticipate future politicization on a wide variety of issues
pertaining to COVID-19. When prompted to speak in the style of a liberal
Democrat, the model exhibits a greater likelihood of choosing to be
vaccinated, choosing to wear a mask, endorsing vaccine and mask
mandates, and supporting lockdowns. By contrast, when prompted to speak
as a conservative Republican, responses are more likely to agree that
wearing a mask or getting vaccinated should be personal decisions rather
than government mandates and that both mask and vaccine mandates should
be ended. To shed light on why the model associates liberals with more
cautious responses than conservatives, we machine code more than 4,500
open-ended justifications for those responses and identify common themes
corresponding to partisan gaps. Specifically, we find that levels of
trust in the government and scientific community as well as the
prioritization of safety versus freedom emerge as common considerations
across questions, grounding novel pandemic-related issues in
longstanding ideological principles.

These findings suggest that certain features of contemporary American
liberalism and conservatism structured the way COVID-19 politicization
unfolded. While these findings do not imply that ``discourse is
destiny,'' they provide compelling evidence that existing ideological
systems channeled the reception to this novel issue in a way that
ultimately undermined a unified public response. Our results suggest
that COVID-19 did not fundamentally alter American political ideology,
but our research design could similarly be applied to identify instances
where history does deviate from the projections based on prior
discourse, suggesting an unanticipated development of discourse. When
social movements or influential public figures cultivate surprising new
discourses that do not conform to existing patterns, this would result
in a disjuncture between prior models and subsequent empirical
observations. The failure of models to anticipate future developments
may therefore suggest points when ``history matters,'' and key events
reshape public discourse in unpredictable ways.

Interviewing simulated respondents with an AI model presents obvious
limitations. Most critically, an LLM may inaccurately reproduce a
discourse, leading to incorrect or misleading inferences
\parencite{Ji2023-ut}.
Nevertheless, in situations where the target population cannot be
interviewed or a question requires measuring responses at a scale that
is practically infeasible, simulated respondents may provide the best
available evidence into such questions that otherwise elude empirical
analysis \parencite{Kim2023-ji, Kozlowski2019-vh}. Investigations with simulated
respondents thus occupy the edge of empirical sociology; the information
they provide may not definitively resolve a debate, but can place
meaningful evidentiary weight on important questions that lie outside
the reach of conventional methods.

\section*{THEORETICAL FRAMEWORK}

Advocates for the use of artificial intelligence in science often
emphasize algorithms' capacity to surpass human performance. Indeed,
such ``superhuman'' abilities are already facilitating important
scientific contributions. In social science, machine learning algorithms
assist researchers by identifying objects in videos and photographs,
transcribing audio into text, and classifying texts into typologies, all
at speed and scale far surpassing human capabilities
\parencite{Bonikowski2022-th, Grimmer2022-fu, Hannan2022-bs, Le_Mens2023-wa, Vicinanza_undated-cj}. In the
natural sciences, algorithms are beginning to make new discoveries by
combining computational power and speed with sophisticated knowledge
bases. Arguably the most important advance has been AlphaFold's success
at solving protein folding \parencite{Senior2020-ee}, but advances in drug \parencite{Jimenez-Luna2020-tg} and materials discovery
\parencite{Wilkins2023-mx, Zhou2018-cw}, the control of complex nuclear fusion reactors
\parencite{Degrave2022-zi}, and
even the identification of novel auction and market policies are also
promising \parencite{Jiao2021-mb, Mosavi2020-hy}.

Social scientists, however, may gain more from artificial intelligence
by capitalizing not on its capacity to surpass human performance, but
its ability to mimic it
\parencite{Brynjolfsson2023-yu, Sourati2023-aa}. Simulation studies in the social sciences have
historically favored elegant models with simplistic agents over
empirically realistic ones. Such formal models provide important insight
into how complex social patterns emerge from simple interactions, but
they tell us comparatively little about the dynamics of specific
empirical cultural systems, organizations, or institutions. Progress on
this front has long been hindered by the difficulty of specifying
empirically realistic agents to populate complex social simulations.
Fortunately, due to their training on massive archives of rich cultural
data, modern AI models can now generate ``digital doubles'' of human
respondents, capable of faithfully reproducing the knowledge,
preferences, and behaviors characteristic of a specific social group.

Social simulations with empirically realistic agents open productive
avenues for research that would be impossible with human subjects.
First, AI models can use textual records to reproduce the discourse of
social groups that no longer exist or that otherwise cannot be
interviewed. Because these models are generative, they enable analysts
to go beyond the exact statements made in the textual archive and
extrapolate likely out-of-sample utterances consistent with the semantic
associations in observed texts. Second, digital doubles can produce
simulated data at scale. Millions of interactions between digital actors
can be quickly and affordably simulated over a wide array of initial
conditions and differently parameterized agents, detailing a richer and
denser high-dimensional interaction space than would be possible with
human actors \parencite{Lai2024-qg}. Lastly, the internal representations of AI agents are directly
observable in a way that human representations are not. Although the
activation patterns of deep neural networks are commonly described as
black boxes, these representations can be directly analyzed and explored
for deeper understanding.

\subsection*{How Large Language Models Work}

Computational models for text analysis have already found widespread
application in the social sciences
\parencite{Gentzkow2019-vu, Grimmer2022-fu}. The most recent innovation to gain
prominence is the word embedding model, which represents semantic
relationships between words in a text as geometric relationships between
word vectors in a high dimensional space
\parencite{Mikolov2013-va,
Pennington2014-dz}. Words that are used in similar
contexts (and therefore share similar meanings) are positioned close
together in the embedding space, whereas words that occupy very
different contexts are located far apart. Social scientists have shown
that the positioning of words in an embedding space preserves cultural
information from model's training texts, such as words' connotations of
masculinity or femininity, affluence or poverty, and thought or action
\parencite{Boutyline2023-sr, Garg2018-jf, Kozlowski2019-vh, Stoltz2019-mx}.

Although word embedding models mark a major advance in learning and
representing semantic relations, they remain ill-suited for the task of
generating new texts. Modern LLMs outperform classical word embeddings
at task of language modeling for three key reasons: (1) LLMs are
autoregressive models that iteratively predict the next word rather than
the central word \parencite{Brown2020-nh}, (2) they use ``self-attention'' to imbue their word vectors
with local contextual information \parencite{Vaswani2017-wi}, and
(3) they leverage a deep neural network architecture that enables the
learning of more varied and complex linguistic patterns \parencite{Kaplan2020-lo}.

\subsubsection*{Autoregressive language modeling}

\begin{figure}
  \captionsetup{justification=raggedright,singlelinecheck=false}
  \caption{Autoregressive Language Modeling as a Branching Tree of Possible Completions}
  \centering
  \includegraphics[width=0.6\textwidth]{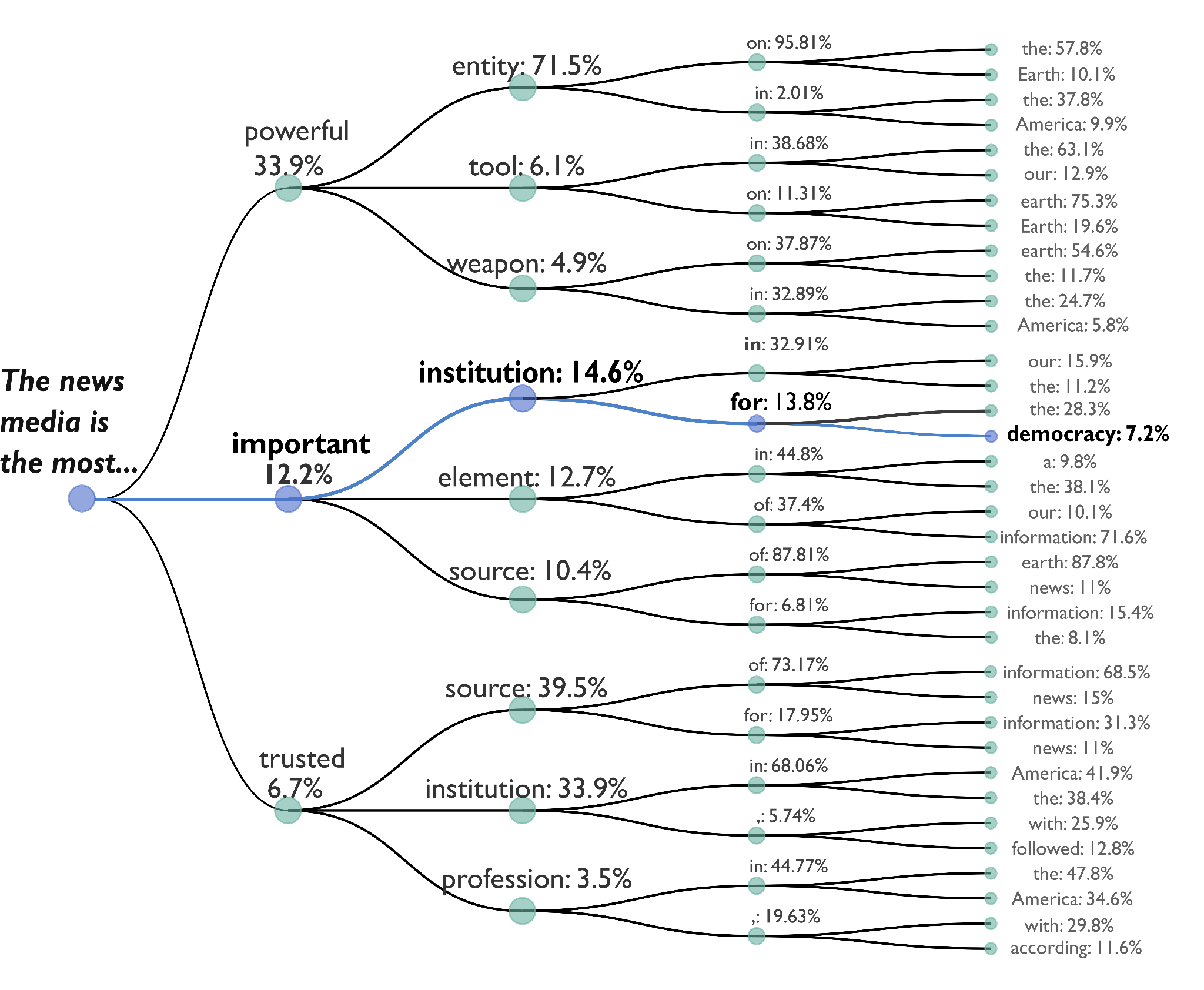}
  \label{fig:autoregress}
\end{figure}

LLMs such as GPT-3 are ``autoregressive'' language models, meaning they
optimize the prediction of the next word given a sequence of previous
words. Such models take as their input a ``prompt,'' and conditional on
the sequence of words comprising the prompt, they generate a probability
distribution for the next word in the sequence. A word is then randomly
drawn according to this probability distribution and is appended to the
original prompt. The next-word prediction task is then repeated using
this newly extended prompt, generating yet another ``next word.''
Because each new word is drawn stochastically from a probability
distribution, we can conceptualize text generation as following a single
pathway through a branching tree of potential next-words, growing a
sentence one word at a time (Figure \ref{fig:autoregress}). By repeatedly inputting the same
prompt to an LLM, an analyst can generate a distribution of directions a
statement is likely to take.\footnote{Some LLMs directly draw upon the
  branching tree of next word probabilities in order to maximize the
  likelihood that the entire response, and not just each word, is most
  probable. By keeping a broad ``beam'' or tree of probable texts, the
  model can wait and select the one more probable at the end. Other
  models, including OpenAI's GPT family, do not use beam search
  explicitly, but rather run self-attention layers many times (so-called
  multi-headed attention) to produce a distribution from which the top
  probability output text can be selected. \hl{}} This autoregressive
approach differs from early word embedding models like word2vec or prior
transformer-based models like BERT which used preceding \emph{and}
following words to predict a central target word
\parencite{Devlin2018-um, Mikolov2013-va}. This bi-directional approach may benefit from the
additional information of subsequent words but is inappropriate for the
task of generating new text, in which only prior words are
available.\footnote{Ilya \citet{Sutskever2023-sa} has
  posited that autoregressive models may outperform bi-directional
  models precisely because their task is more challenging and therefore
  induces more thorough learning.}

Equation \ref{eq:autoregressive} describes how the autoregressive language models calculate the probability of a given sequence of words $(y_1, y_2, \ldots y_n)$ as the product of the probabilities of each word $(y_t)$ conditional on the prior words in the sequence $(y_{<t})$.

\begin{equation}
  P(y_1, y_2\dots y_n) = \prod_{t = 1}^{n}p(y_t|y_{<t})
  \label{eq:autoregressive}
\end{equation}

Given the multiplicative nature of joint probabilities, any statement
more than a few words long tends to exhibit a very low probability.
For example, the most likely completion in Figure \ref{fig:autoregress},
``\emph{powerful entity on the}'', would occur with a probability of 13.4\%.
(0.339 * 0.715 * 0.958 * 0.578). The completion we highlight, ``\emph{important
institution for democracy'',} is one of the most likely four-word completions,
yet it occurs with a probability of less than 0.1\%. Extending to longer
generations, the probability of any given completion becomes smaller yet.
Joint conditional probabilities thus produce a branching structure that spans
a vast diversity of completions, each one with a low individual probability of
occurring.

It is important to note that autoregressive language modeling is not a
new concept. Attempts at implementing this approach began as early as
the 1910s with Andrey Markov
\parencite{Gagniuc2017-xm, Hayes2013-yb} and continued into the 1940s with the work of Claude Shannon
\parencite{Shannon1951-ct}, finally
maturing into full statistical models of future word prediction by IBM
in the 1980s \parencite{Rosenfeld2000-rq}. But despite their theoretical promise, probabilistic language
models failed to consistently produce diverse, meaningful, and
grammatically correct sentences for many decades. It was only with two
further advances -- self-attention and deep neural networks -- that
autoregressive language modeling finally achieved success.\footnote{Another
  advance central to modern post-ChatGPT language models is the
  deployment of user feedback in the tuning of the models for human
  interaction and simulation. By fine-tuning towards human responses,
  such models effectively return to the sequence-to-sequence (Seq2Seq)
  architecture, critical in translational models
  \parencite{Xue2020-qb} and
  early transformers \parencite{Vaswani2017-wi}, that considers at each word not only the word that came
  before, but also the priming motivation or ``prompt'' (\emph{x}), be
  it a modeled human response, a picture to be captioned, a turn in
  online conversation, or anything else that inspires the generation of
  text \parencite{Ouyang2022-xw}.}

\subsubsection*{Self-Attention}

LLMs use distributed vector representations to encode the relations
between words, but they advance beyond classical word embedding models
by incorporating a mechanism known as \emph{self-attention} that imbues
word vectors with information from their local context
\parencite{Vaswani2017-wi}.
During training, word embedding models like word2vec treat local context
as a ``bag of words,'' ignoring sequence and multi-word interactions
\parencite{Mikolov2013-va, Pennington2014-dz}. Post-training, each word in a word embedding
has a singular vector representation based on its contexts across the
entire training corpus. By contrast, when a prompt is input into an LLM,
self-attention mechanisms share information between all words in the
sequence. Words have a singular representation only in the first layer
of the model; as the sequence progresses through the model, each word's
vector representation is adjusted and ``contextualized'' by the other
words in the sequence.

\begin{figure}[p]
  \captionsetup{justification=raggedright,singlelinecheck=false}
  \caption{Schematic Diagram of Self-Attention.}
  \centering
  \includegraphics[width=\textwidth, height=\textheight, keepaspectratio]{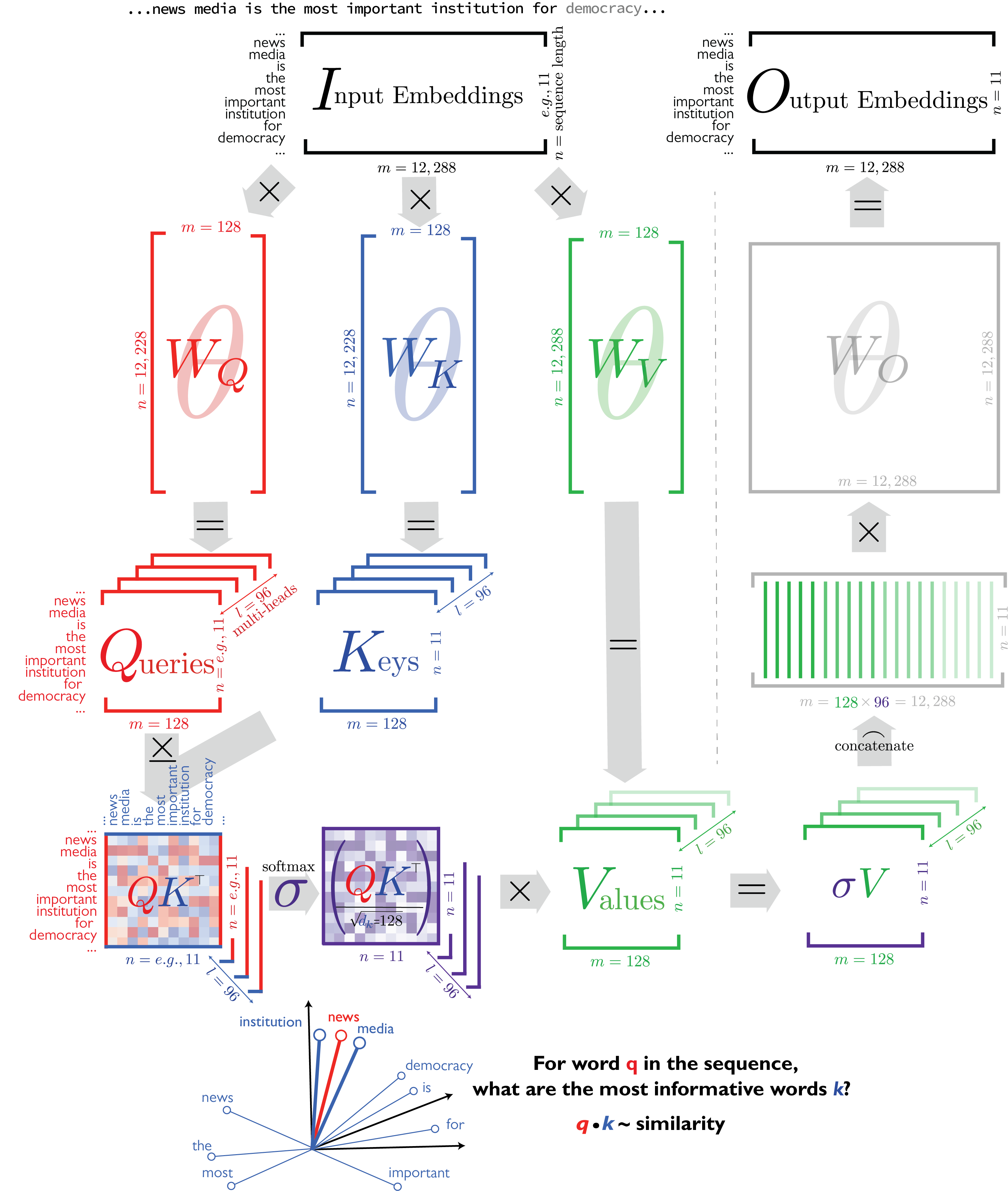}
  \label{fig:schematic1}
\end{figure}

A schematic overview of the self-attention mechanism is displayed in
Figure \ref{fig:schematic1}. Input embeddings are transformed through multiplication with
learned weight matrices (W\textsubscript{K} ,W\textsubscript{Q} ,
W\textsubscript{V}) and subsequent multiplication between the resultant
Key, Query, and Value matrices. Prior to multiplication, however, each
weight matrix is split into many smaller matrices (96 in GPT-3) by
column in what is called ``multi-head attention.'' Multi-head attention
enables the model to attend to multiple versions of the input sequence
simultaneously while also easing computation with improved
parallelization. After the input embeddings are multiplied by the weight
matrices, the key and query matrices are themselves multiplied, creating
a square $n * n$ matrix for a prompt of length \emph{n}. Each entry
{[}\emph{n\textsubscript{i}},\emph{n\textsubscript{j}}{]} in this square
matrix captures a meaningful ``interaction'' between token \emph{i} and
token \emph{j} in the prompt that answers the question ``for query token
\emph{x}, what key tokens \emph{y} from the sequence provide the most
informative context''. These attention weights, the dot products of
\emph{x} and \emph{y}, are then normalized by the square root of the Key
dimension and transformed via softmax, the multiple-outcome
generalization of the logistic function, so that each row's
values sum to 1 and the predictive power of each key token on a query
token can be interpreted as a probability. These probability estimates
are treated as weights and reshaped through multiplication with the
``value'' vectors for each token. The heads of the weighted Value matrix
are then concatenated back into a single wide matrix which is multiplied
by a final output weight matrix (W\textsubscript{O}), producing the
output embeddings that pass through a feed-forward neural network before
exiting the transformer block. Across this self-attention layer, the
learned parameters include the weight matrices (indicated with a $\theta$),
and the input embeddings for the very first model layer.

\FloatBarrier
\FloatBarrier

This complex operation has three important implications. First,
self-attention solves the problem of polysemy. In classical word
embedding models, the multiple meanings of words are conflated within a
single vector representation. Self-attention resolves this issue by
adjusting the position of each word vector using information from all
other words in the prompt. For instance, the word ``bark'' would be
modified in one way if it is referencing a tree, and another if
referencing a dog. Self-attention similarly resolves anaphora, linking
pronouns with their associated nouns by sharing semantic information
between them. Although pronouns like ``it'' carry little information in
a classical word embedding, they can be meaningful when linked to the
correct noun by self-attention in an LLM. Lastly, self-attention can
up-weight words most relevant for predicting the next word. For example,
when predicting the next word in the sentence, ``Benjamin Franklin,
noted inventor and statesman, was born in the year \_\_\_\_\_\_'' the
words ``Benjamin'' and ``Franklin'' (and the interaction between them)
are more important in predicting the next word than terms such as
``inventor'' or ``noted.'' These highly relevant words may therefore be
up-weighted for predicting the next word. Thus, even long-range
dependencies can be preserved and leveraged by ``contextualizing'' and
sharing information between words (Figure \ref{fig:schematic1}).

Self-attention highlights another key difference between LLMs and
earlier models such as word embeddings -- \emph{a} \emph{single LLM is
able to preserve a variety of distinct discourses} \parencite{Argyle2023-ii}. In a
word embedding, each word has a single relationship to every other word,
represented as proximity between the respective word vectors. To compare
two discourses using word embeddings, an analyst would need to train
independent models on separate collections of text representing the
desired discourses, then compare the relative positioning of word
vectors between models. By contrast, words in an LLM do not have a
single representation; each word's vector representation is modulated by
the presence and order of other words in the prompt. Thus, if a word or
phrase has a different usage across discourses, these multiple senses
can be preserved in a single model and activated by surrounding context
words. This means that many different discourses can be generated with a
single LLM by inputting prompts that prime different cultural registers,
so long as the training corpus includes sufficient texts to learn the
linguistic patterns of the respective group
\parencite{Argyle2023-ii}.

\subsubsection*{Deep Architecture}

The final factor enabling LLMs' success in producing humanlike texts is
their massive neural architecture. Word embedding models like word2vec
use a shallow architecture with a single hidden layer, and the total
number of parameters learned by the model is typically in the tens of
millions.\footnote{The number of parameters in a word2vec model is the
  product of vocabulary size (N) and the user-specified number of
  dimensions (M) multiplied by two, because the model simultaneously
  learns the hidden ``context embedding'' along with the final ``word
  embedding,'' each of size NxM. A typical word2vec model may have a
  vocabulary of size 50,000 and 300 dimensions, resulting in 30 million
  parameters.} By contrast, GPT-3's neural network consists of 96 layers
and 175 billion parameters
\parencite{Brown2020-nh}. GPT-4's
architecture has not been formally released, but expert consensus is
that GPT-4 is substantially larger, with parameters likely numbering
over one trillion \parencite{OpenAI2023-kd}. A greater number of parameters and layers enables a neural
network to learn more complex functions. For a language model to
faithfully encode the multitude of discourses that appear on the
internet, it must compose an exceptionally complex function. This
function leverages the extensive non-linearities and interactions
between words to transform an input sequence into an accurate
probability distribution for the next word. In learning these complex
patterns of linguistic entailments from its training texts, the model
effectively learns the internal structure of a discourse.

\begin{figure}[!htbp]
  \captionsetup{justification=raggedright,singlelinecheck=false}
  \caption{Schematic Diagram of GPT-3 Architecture, from Input Embeddings through Token Prediction.}
  \centering
\includegraphics[width=0.75\textwidth]{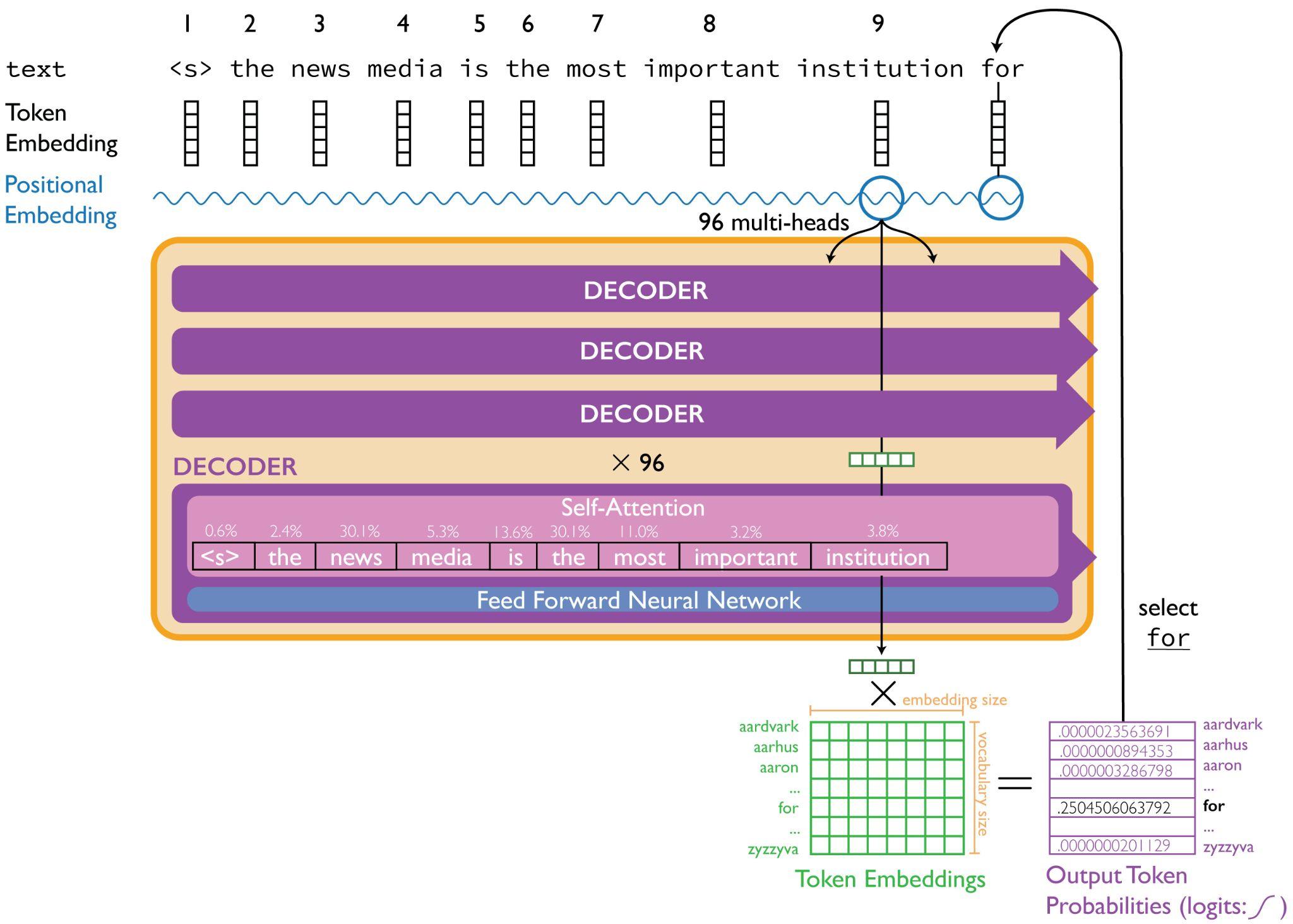}
\label{fig:schematic2}
\end{figure}

Figure \ref{fig:schematic2} presents a schematic diagram of GPT-3's architecture. During
training, a sliding window of words from the training text is used as
the context, and is represented as a matrix of corresponding word
vectors. This collection of word vectors then passes through 96
``transformer blocks.'' Each transformer block comprises attention
mechanisms that are themselves divided into 96 ``attention heads''
followed by a feed-forward neural network which further transforms the
contextualized word vectors in preparation for predicting the next word.
After passing through all transformer blocks, the resulting matrix is
converted into a single vector corresponding to the model vocabulary.
This vector is rescaled into a probability distribution via the softmax
function, a multi-category generalization of the logistic curve. This
probability distribution is then compared to the correct response -- the
actual next word in training texts. Error, or ``loss,'' is calculated as
a function of the difference between the predicted probabilities and the
correct next word,\footnote{GPT-3 minimizes cross-entropy loss.} and
this error is propagated back through the model to update parameters
such that the same prediction would be more accurate if made again. The
context window then continues its progression through the training
texts, repeating the task of predicting each subsequent word given the
prior words. The algorithm may iterate over the entire corpus of
training texts multiple times until improvements become negligible and
training is halted.

Once fully trained, LLMs can generate novel texts by the same operation
of next-word prediction that is optimized during training. The user
feeds the model a prompt and the algorithm iteratively predicts the
following words one by one, appending each newly generated word to the
prompt to predict the subsequent word. In this way, the LLM builds new
sentences word-by-word. The primary difference between training and
generation is that there is no ``correct next word'' during text
generation, but both tasks are fundamentally grounded in the
next-word-prediction task.

Prominent chatbots like ChatGPT use an LLM as their foundation, but are
then adjusted by their designers to respond to queries in the style of a
helpful assistant. This tuning process, known as Reinforcement Learning
with Human Feedback (RLHF), steers the model toward responding to
queries rather than simply predicting the next word in a sequence, while
also censoring content that is offensive or controversial
\parencite{Ouyang2022-xw,}. The
original GPT-3, however, was released prior to fine-tuning with RLHF,
and therefore simply predicts the next word in the user-provided
sequence without additional weighting or censorship. This makes GPT-3 a
more straightforward tool for the recovery and analysis of historical
discourses, although recovering a variety of discourses from models
fine-tuned with RLHF is also possible.

Using a basic LLM such as GPT-3, the user can simulate responses from a
socially-situated respondent by prompting a model with a statement that
primes a given perspective. For instance, by starting a statement with
``I am a conservative Republican and I believe'' will produce
systematically different completions than a statement that starts ``I am
a liberal Democrat and I believe''
\parencite{Argyle2023-ii}. The
differences in the completion will reflect the differing semantic
associations of ``conservative Republican'' and ``liberal Democrat''
learned from the model's training texts. Leveraging information from a
near complete record of text from the internet, the model will attempt
to generate likely endings for each of these phrases. This technique
thus presents a novel means of interrogating the system of claims,
considerations, and justifications that characterize an ideology,
extending far beyond the simple positions of words obtained from a
classical word embedding model.

The process of generating text in response to a prompt with a
pre-trained model is called ``inference'' in computer science. Recent
work demonstrates how the inference process, as described above,
directly optimizes text responses to user prompts to maximize syntactic,
semantic, and even pragmatic appropriateness. This work demonstrates
that the number of transformer layers in contemporary transformers
(e.g., 96 for GPT-3) is comparable to the number of steps required to
optimize neural network weights, resulting in an optimized and
efficiently produced textual response that does not require additional
neural network training \parencite{Von_Oswald2023-xc}. Related work demonstrates how inference from
prompts unleashes in a within-model process of gradient descent for
error minimization akin to the process of fine-tuning a model from
external data \parencite{Dai2023-bg}. In sum, transformers enable the potential to model multiple,
well-defined human perspectives efficiently in ways that both enable
analysis and generative simulation.

\subsection*{Digital Doubles and the Study of Ideology}

Through their ability to recreate discourse models underlying their
training texts, LLMs offer new avenues for studying cultural-historical
change and advancing theories of meaning. In particular, we use LLMs to
shed new light on the nature and extent of \emph{constraint} in cultural
systems. Structuralist theories emphasize the overall coherence of
meaning systems, positing that seemingly elaborate systems of
classification and evaluation are reducible to simple underlying logics
\parencite{Douglas1966-re, Levi-Strauss1966-fw}. The diverse array of practices, values, and beliefs
expressed within a culture can be understood as ``all of a piece,''
unified by subtle threads of configured meaning that can be revealed
through careful analysis
\parencite{Mead1942-zm}. This coherent
model of collective meaning has important implications for cultural
change; new ideas can only be integrated into an ideology if consistent
with the overarching logic of the system. By this theoretical model,
cultural evolution is largely predictable because it is highly
constrained. When a new object enters a cultural context, its potential
interpretations are tightly limited by the logic of the cultural
context.

This constrained model of cultural coherence was largely displaced by a
wave of scholarship that emphasized the importance of historical
contingency and internal contradictions of cultural systems. To redress
the structuralists' failures to capture the apparent arbitrariness and
contradiction of culture, the image of a unified system was supplanted
within sociology by a model of culture as a ``toolkit,'' a repertoire of
strategies that can be taken up or discarded as necessary
\parencite{Swidler2003-il}, or a largely
disconnected set of cognitive schemata which are deployed situationally
\parencite{DiMaggio1997-rx} to satisfy a
pragmatic purpose. More recent work has sought to identify a theoretical
middle-ground, accepting that systems of meanings exhibit broad
patterning, but remain only weakly constrained, rife with instability,
ambivalence, and contradiction
\parencite{Baldassarri2014-mi, Boutyline2017-vj, Kiley2020-hi, Rawlings2020-yb, Swidler2003-il}.

A similar debate has unfolded in the field of political opinion. One
strand of research argues that voters hold core values and form opinions
on specific issues in accordance with basic underlying considerations
such as freedom, equality, or protection from harm
\parencite{Feldman1992-lf, Goren2016-rw, Haidt2012-fy, Lakoff2010-yy}. This
theoretical model posits that voters' conceptual systems are constrained
by an internal logic with a few core values structuring a complex
attitude system that covers a diverse array of specific issues.
Detractors of this theory argue that individuals' attitudes are steered
not by internal values, however, but by partisanship. According to this
top-down alternative, partisan leaders construct a platform of positions
through a strategic process of ``log-rolling,'' with the aim of building
a coalition across various interests
\parencite{Carmines1989-yi, Zaller1992-eq}. Partisan leaders then broadcast this assortment of
positions as a unified platform, which is then absorbed \emph{in toto}
by strong partisans in the electorate
\parencite{Green2004-eq}. By this model, the collection of opinions that
constitute an ideology are the arbitrary result of historical-political
contingencies; held together not by a unifying cultural logic, but
social and political messaging.

This debate over ideological coherence revolves around a key empirical
question. When a new issue emerges, are public responses already
prefigured by ideology, or is a new issue ideologically indeterminate
until partisan elites voice positions on the issue
\parencite{Noel2014-es, Page2010-yj, Zaller1992-eq}? If political ideologies are general
dispositions that inform attitudes across issues, they could be readily
transposed to new issues without elite direction. If ideologies are
arbitrary assortments of attitudes strategically drawn together through
political coalitions, however, then the politicization of a new issue
would be unpredictable until partisan elites broadcast their positions.
Thus, study of the exogenous injection of a new issue within a political
landscape sheds light on a key theoretical question in the sociology of
culture -- when a new idea is introduced into an existing system of
beliefs, is it constrained by the logic of the belief system or free to
take any direction?

LLMs offer a powerful new approach for exploring this core question in
the study of culture and ideology. Typically, by the time the public
learns about a new issue, it has already been politicized by partisan
elites. Thus, when ideology in the electorate mirrors party platforms,
analysts are unable to adjudicate if elite cues steered public opinion
or if underlying ideological convictions shaped the new issue's
interpretation for both party elites and the public at large. The key
benefit of LLMs is that they capture and preserve discourses from the
historical period of their training texts. Therefore, an LLM trained on
texts from a given time can simulate respondents from that period and
present them with questions that anticipate future cultural or political
developments. While previous text analytic models produce
representations of cultural systems from historic texts, LLMs are
\emph{generative} models to which researchers can pose novel questions.
These prompts could even include hypothetical scenarios or issues that
only emerged after the model's training. By leveraging the vast
linguistic information learned in training, generative cultural models
produce the most likely responses to novel questions given the
discursive associations of that period. This method enables new insight
into the historical process of cultural change, providing a lens through
which we can assess which developments are truly surprising and which
are already prefigured by the cultural system.

Moreover, the generative nature of LLMs enable us not only to assess
distinct perspectives, but to interrogate them. In his classic 1940
article on ``Situated Actions and Vocabularies of Motive'', sociologist
C. Wright Mills argued for the importance of analyzing the shared
language by which persons from distinct socio-cultural situations
justify and account for their positions and conduct
\parencite{Mills1940-xp}. This was not
fixed on the psychology of human motivations and drives but rather the
sociology of how certain motives are more acceptable given certain
audiences and contexts than others. But in addition to being
situationally defined, social appropriateness of a motive is also
constrained by identity. As Goffman\emph{'s} dramaturgical perspective
suggests,``when an individual plays a part he implicitly requests his
observers to take seriously the impression that is fostered before them.
They are asked to believe that the character they see actually possesses
the attributes he appears to possess, {[}and{]} that the task he
performs will have the consequences that are implicitly claimed for
it....'' \parencite{Goffman2021-xf}. To the extent that an LLM encodes the discursive patterns of a
social group, the analysts can go beyond simply ``polling'' the
simulated respondents. They can prompt the model to justify its
responses within a designated cultural register, revealing a set of
linked considerations that render the expressed attitude coherent.

Mills' vocabulary of motive, however, does not reach its own aspiration.
A ``vocabulary'' implies a collection of low-level words that might be
flexibly deployed according to some higher-order cultural logic to
plausibly and acceptably justify a position. It implies the prevalence
of templated excuses that could be tallied and expected in particular
contexts. Modern LLMs allow us to reach past this conception to model
not only situated vocabularies of motive, but their syntax and
semantics. LLMs can generate and discriminate how even novel, unique
expressions of motive should be more or less expected from a situated
actor regarding their behavior. How can we do this with LLMs? As with
people: By asking them over and over again.

\subsection*{2019 Politics in Silico}

The spread of COVID-19 to the United States presented a critical event
for social and political meaning making. Lacking any precedent in living
memory, the pandemic was not readily interpretable within existing
frames for political response. This is evident from the earliest public
opinion surveys, which show relatively little partisan division on
questions relating to the virus
\parencite{Deane_undated-zp}. Nevertheless, political elites and opinion leaders began to
broadcast a variety of competing interpretations of the situation as
soon as a virus was detected in the US in January of 2020
\parencite{Stokes2020-vw}. By
March 2020, a sizable divide had already grown between self-identified
democrats and republicans regarding the appropriate government response
to the emerging pandemic
\parencite{Gadarian2021-su}. This
politicization of COVID-19 would prove to be a defining characteristic
of the pandemic period, imbuing discussions of lockdowns, masks, and
vaccines with partisan fervor, ultimately stymying any unified public
response to the virus
\parencite{Albrecht2022-kj, Allcott2020-iy, Chen2022-eb}.

The politicization of COVID-19 related issues quickly assumed a common
pattern. Liberals viewed the virus as an urgent threat warranting
immediate and sweeping response, whereas conservatives questioned the
danger posed by the virus, denounced responses that infringed on
individual liberties, and doubted the safety of the
government-sanctioned vaccine. After years of pandemic politics, this
familiar pattern may appear self-evident, and there are indeed some
reasons to view this polarization as predictable. On a variety of
issues, ranging from gun control to universal healthcare to motorcycle
helmet requirements, liberals tend to favor protection and government
intervention while conservatives lean towards freedom and personal
choice \parencite{Homer2009-gw}. To explain this pattern, some public opinion analysts have
argued that conservatives more highly value freedom while liberals
prioritize considerations of equity and protection from harm \parencite{Feldman1992-lf, Haidt2012-fy}.

Nevertheless, a large literature from political psychology plausibly
anticipates the very opposite empirical outcome and supplies numerous
reasons to expect that conservatives would support stricter responses to
the virus than liberals. In an influential review of the psychological
correlates of political ideology, Jost
\parencite{Jost2006-wm} cites robust
international evidence that political conservatism is associated with
(i) fear of death, (ii) aversion to threat or loss, (iii) uncertainty
avoidance; and (iv) needs for order, structure, and closure. Each of
these predispositions suggest that \emph{conservatives} should be the
ones to advocate for harsher measures to protect against the virus, not
liberals. Moreover, a wide array of studies suggest that conservatism is
associated with desire for purity and aversion to contamination. This
literature emphasizes forms of social or religious impurities, but more
broadly connects political conservatism to a general fear of
contamination and uncleanliness
\parencite{Haidt2012-fy, Helzer2011-ni, Jost2017-sj, Oxley2008-pi, Terrizzi2013-po}. Consistent with this pattern, Republicans
were more concerned about the Ebola epidemic of 2014 than Democrats
\parencite{Pew_Research_Center2014-gi}.

Similarly, there are reasons to suspect that liberals would be skeptical
of sweeping government responses to the virus. Liberals have a long
history of skepticism toward vaccines, arguing that they are unnatural
and pushed by large, profit-driven pharmaceutical companies
\parencite{Callaghan2019-zh, Colgrove2006-jy, Conis2014-qi, Jamison2019-yo}. Also, in recent historical cases where the safety of the
American people has been at stake, such as the War on Terror,
conservatives were more willing to sacrifice personal liberties for
public safety than liberals
\parencite{Rosentiel2011-pi}. All of
these considerations suggest that history could have played out
differently, and that an alternate framing for COVID-19 was plausible,
which would steer conservatives to endorse cautious measures and
liberals to oppose them.

It is possible that prior values and predispositions do little to steer
the public response to an issue; any political topic could be framed in
a variety of ways, and predispositions provide little direction until
partisan elites provide an interpretation of the issue in terms of clear
ideological considerations. The alternative hypothesis is that ideology
does steer reactions to a new issue like COVID-19. Even if the initial
effect of ideology was weak, it could set off a cascade of
self-reinforcing dynamics among both elites and the public, ultimately
setting the course for widespread polarization
\parencite{DellaPosta2020-ta, Rawlings2022-pw}

Assessing whether the public was inclined to tip towards a given form of
politicization prior to elite signaling would require measuring public
response to pandemic-related questions \emph{before its top-down
framing}. In practice, this is difficult because most issues are rapidly
framed by elites before social scientists can measure public attitudes.
Because collecting public attitudes to an issue before it emerges is
unfeasible, the best alternative is to reconstruct the discursive space
prior to the emergence of a novel issue and interrogate its associations
surrounding the issue. We therefore use GPT-3, the first LLM capable of
faithfully reproducing complex attitude systems, to reconstruct the
political landscape of 2019 so we may investigate whether this period
exhibited a predisposition to tip towards the pattern of polarization
that manifested in the following years.

\section*{DATA AND METHODS}

For the following analyses, we analyze completions generated by the
GPT-3 language model. This approach presents a notable departure from
conventional methods of text analysis. LLMs are generative models, and
the most straightforward way to learn from these models is not to
examine their internal representations but to study the outputs they
produce. As a result, the data we analyze are not actual statements made
by members of our target populations nor are they representations of
such statements like topic models or word embeddings. Our data are novel
word sequences learned to occur with high probability given the
discursive patterns learned in a vast training corpus.

Such an approach deviates from the simulation studies typical of formal
sociology. Formal models commonly attempt to parsimoniously capture
social dynamics by precisely specifying minimal conditions under which
empirically-observed patterns can be reproduced. Using LLM outputs as
data, by contrast, constitutes a hybrid of empirical and formal
sociology; we analyze simulated data, but the simulation is trained to
produce outputs that closely approximate empirical
distributions.\footnote{Previous social simulation studies have used
  parameters estimated by empirical models to improve correspondence to
  observed contexts, but these models still typically strove to
  parsimoniously reproduce the system of interacting factors,
  prioritizing interpretability over predictive accuracy.} To the extent
that the model successfully reproduces a population's response
distributions, the model's outputs can stand in for human responses from
that population and be analyzed at scale. We adopt this approach to
simulate the opinions of American liberals and conservatives in October
of 2019, the historic point immediately prior to the emergence of
COVID-19.

\subsubsection*{Prompt Design}

Our primary aim is to discern whether a speaker identifying as liberal
is predicted to have different views regarding pandemic responses than a
speaker identifying as conservative. We therefore design prompts that
isolate the effect of partisan identification words on the generation of
responses toward a variety of COVID-19 issues. We use three modes of
partisan identification: ideological identification (liberal or
conservative), party identification (Democrat or Republican), and
candidate preference (Hillary Clinton or Donald Trump). To maximize the
partisanship signal, we use all three of these identifiers in all our
prompts.\footnote{Robustness tests suggest that removing the candidate
  preference from prompts generally weakens effects but rarely
  eliminates them.} All prompts begin with a ``partisan priming'' taking
the following form:

\emph{``I am a strong \textbf{conservative} and a lifelong
\textbf{Republican}. In 2016, I was proud to vote for \textbf{Donald
Trump} and I think that the \textbf{Democrats} have been a disaster for
this country.}

or, conversely:

\emph{``I am a strong \textbf{liberal} and a lifelong \textbf{Democrat}.
In 2016, I was proud to vote for \textbf{Hillary Clinton} and I think
that the \textbf{Republicans} have been a disaster for this country.}

Because GPT-3 was trained on texts published only through October 2019,
it has no knowledge of COVID-19. This ignorance is an asset, as it
allows us to investigate the ideological landscape immediately prior to
the emergence of this pivotal issue. However, in order for the model to
produce an informative response, we must supply some basic knowledge of
the virus to the model in the prompt. Directly following the initial
``partisan priming,'' we insert the following sentence:

\emph{``Lately, one of the biggest political issues has been the
COVID-19 pandemic caused by the new coronavirus. There is a lot of
controversy around \{issue\}.}

We replace the \{issue\} variable with 49 different issues relating to
different aspects of COVID-19, such as ``whether wearing face masks in
public places should be optional or mandatory'' or ``whether students
should be required to receive the COVID-19 vaccine before returning to
school.''\footnote{Exact wordings of the 49 issues are provided in the
  Appendix.} Within each issue, we test a variety of question wordings.
Some variations are designed to test whether certain key words steer
responses (e.g. vaccine \emph{mandate} vs. vaccine \emph{requirement}),
and other variations are included simply to improve robustness (e.g.
``\emph{I think} face masks are'' versus ``\emph{In my opinion}, face
masks are''). Conventional psychometric surveys are limited by the cost
of additional questions and the limits of respondents' attention
\parencite{Furr2021-ud}. However, with
relatively inexpensive digital doubles, we can relax these constraints
and explore the sensitivity of answers across a range of differently
worded prompts.

\begin{figure}[!htbp]
  \captionsetup{justification=raggedright,singlelinecheck=false}
  \caption{Outline of the prompt design process and conversion of open-ended responses to scores on a semantic axis defined by two anchoring terms (e.g., ``bad idea'' and ``good idea'')}
  \centering
\includegraphics[width=0.8\textwidth]{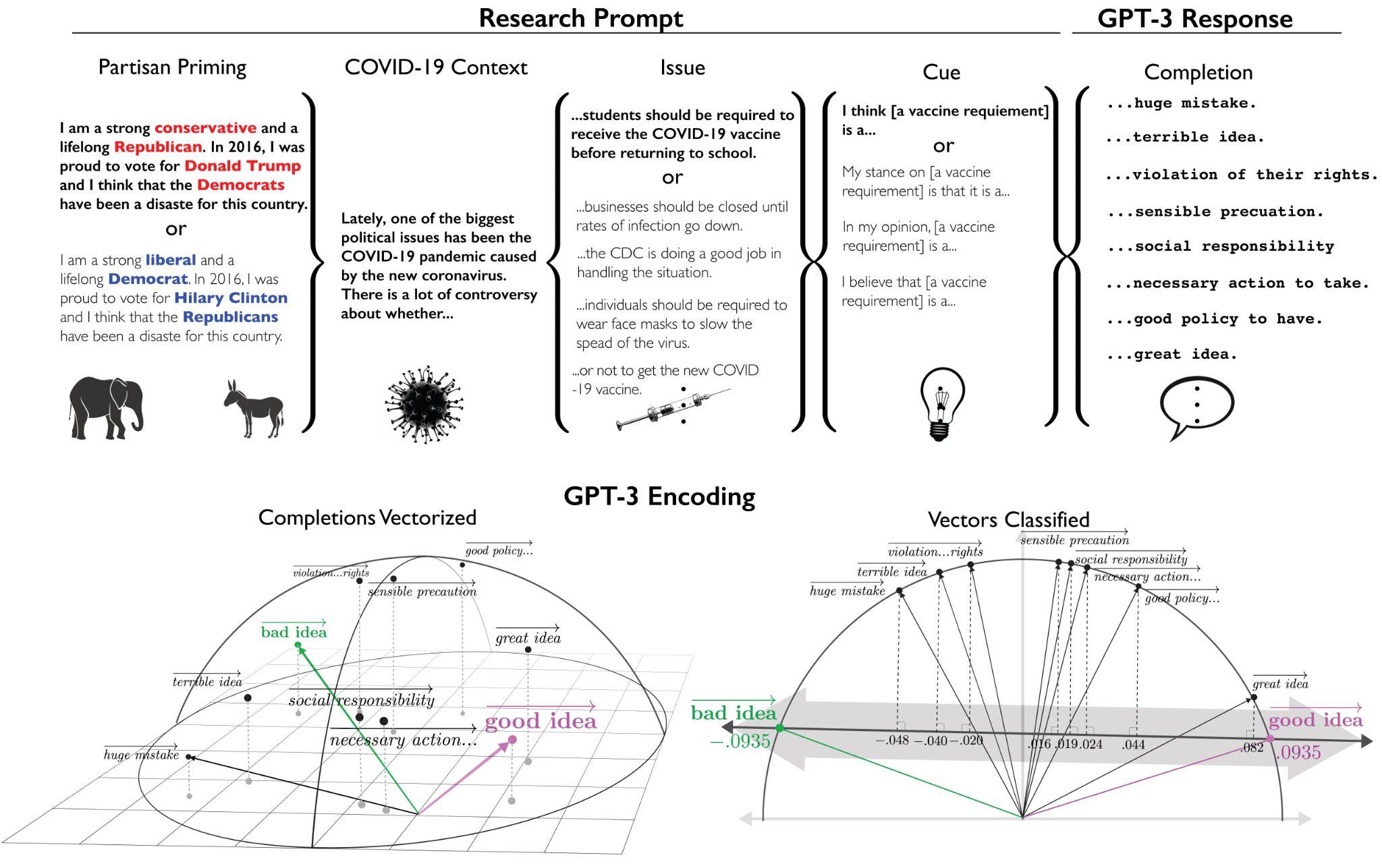}
\label{fig:promptdesign}
\end{figure}

After introducing the issue, the prompt ends with a statement such as
``I believe this is a'', after which GPT-3 begins its completion. In
pilot testing, we found that the more open-ended ``I believe'' led to
many non-committal outputs like ``that some people don't like to talk
about this'' whereas the ending with ``\emph{this is a''} encourages
more clearly valenced responses like ``terrible idea'' or ``great
plan.'' An outline of our piecewise approach to prompt construction in
Figure \ref{fig:promptdesign}.

Including wording variations, our total number of questions reaches 179.
We test each prompt with both a liberal and conservative partisan
priming, doubling the number of unique prompts. We therefore present the
model with a total of 179 x 2 = 358 distinct prompts. To generate a
sample of completions, we input each prompt into GPT-3 500 times,
resulting in a total of 179,000 completions.

\subsubsection*{Machine Coding Responses}

We designed our prompts to encourage relatively standardized completions
indicating either a positive or negative response. Total standardization
was not desirable, however. LLMs are built to be ``programmed'' natively
with language inputs and to produce natural language outputs.\footnote{It
  is possible to make an LLM respond to closed-form multiple choice
  questions, but in order to optimize a model to perform on this task
  the analyst would append an output layer with a softmax activation
  that predicts the best response category rather than the next word.}
In pilot testing, we experimented with prompting the model to produce
closed-form responses, but these structured responses performed markedly
worse than open-ended responses at predicting partisan stances on
well-established political issues. Indeed, open-form responses are
generally more informative than constrained closed-form responses for
human respondents as well
\parencite{Tourangeau2000-tk, Willis2004-lr}. The open-ended responses generated by the
LLM have clear valences, but still take on a wide variety of unique
forms. Given the impracticality of hand-coding 179,000 responses, we use
OpenAI word embeddings to classify completions as either positive or
negative \parencite{Neelakantan2022-yd}. These embeddings are built atop an LLM architecture similar to
GPT-3, but fine-tuned with a contrastive loss that directly pulls
similar words together and pushes dissimilar ones apart
\parencite{Neelakantan2022-yd}.

OpenAI recommends a technique for classification with word embeddings
similar to Kozlowski, Taddy, and Evans' (2019) method for projecting
words onto axes of cultural meaning. We first produce a vector
representation of each prompt completion using GPT-3 embeddings. We use
the ada-text-embeddings-002, which are 1026 dimensional and perform well
on semantic similarity tasks. As with word2vec, words or phrases that
are semantically similar are proximal in the embedding space. Thus, to
classify whether a completion is semantically closer to ``good idea'' or
``bad idea,'' we calculate the cosine similarity between the
completion's vector and the vector for each of those two anchoring
phrases. ``Good idea'' and ``bad idea,'' serve as our anchoring phrases
for most completions, but depending on issue wording they also take
forms like ``mandatory'' and ``optional,'' or ``effective'' and
``ineffective.''\footnote{For some completions, we found that responses
  could take multiple positive or negative forms. For these, we average
  together two classification terms (good idea + personal choice; bad
  idea + public health issue).} After calculating each completion's
cosine similarity to each of the two relevant classification terms, we
calculate the difference between these proximities. The result is a
score between -1 and 1, indicating whether the response is more
semantically similar to the first classification option (e.g. ``good
idea'') or the second option (e.g. ``bad idea'').

We use these scores to test for each issue whether prompts with a
liberal priming are statistically distinct from those with a
conservative priming. For each question wording, we fit an OLS
regression of partisan priming predicting the completion's
classification score. These models reveal whether the speaker
identifying as a liberal is more positive about the issue at hand than
the speaker identifying as conservative.

We ultimately consider GPT-3's forecast to be correct if the effect of
partisanship in the OLS model is in the same direction as a partisan gap
observed with surveys in 2020. We draw upon \citet{Gadarian2021-su} and
published results from the Gallup Panel survey
\parencite{McCarthy2023-xj} to source
``ground truth'' partisan gaps against which we compare our simulated
response distributions.

\subsubsection*{Generating Justifications}

For select prompts, we go beyond simply identifying differences in
responses to COVID-19 and attempt to shed light on why the model is
predicting these differences. We elicit this by prompting the model to
\emph{produce a second sentence justifying its initial response} and
reveal the characteristics of ideologically consistent motives. For
these tests, we input the original prompt to GPT-3 along with the
previously generated completion, which is restricted to one sentence. We
then extend this prompt by beginning a new sentence offering a
justification for the earlier response. Specifically, we append the
phrase ``This is because'' to the end of the prompt to induce a
justification response. We find that alternate wordings produce
substantially similar outputs and include examples in the Appendix.

For each prompt, we generate three ``justification'' completions.
Because we initially generate 500 liberal and 500 conservative responses
to each question, this results in a total of 3000 justifications. To
classify these numerous open-ended responses into a few informative
categories, we again rely on machine coding. As above, we use GPT-3
embeddings to represent each justification as a 1026 element vector. But
because we want categories of justifications to emerge inductively, we
use \emph{k}-means clustering to identify thematic groups instead of
rating responses along a predetermined axis. Because justifications in
favor of a given policy should be qualitatively different from those
opposing the policy, we conduct \emph{k}-means clustering separately for
positive and negative responses as scored in the prior step; for
instance, we first perform cluster analysis on the justifications of
statements \emph{in favor} of mask mandates, then we conduct another
independent cluster analysis of all justifications for statements
\emph{opposing} them. We manually select the number of clusters by
considering three metrics (Silhouette, Calinski--Harabasz, and
Davies-Bouldin scores) along with our qualitative assessment of
interpretability and parsimony. After dividing justifications into
clusters, we generate labels for each cluster using GPT-4. We feed a
random sample of 100 entries from each cluster into a GPT-4 prompt with
instructions to provide labels for each set of responses that describes
their distinctive semantic characteristics.\footnote{We designed this
  prompt to generate cluster labels: ``\emph{The following are clusters
  of semantically similar responses to the question of whether
  {[}issue{]}}. \emph{{[}List of sample texts{]} Please write concise,
  specific, and not overly broad labels for each of the clusters that
  describe their unique theme and distinguish them from the other
  clusters. It does not have to encompass all responses but should
  instead reflect the primary theme evident in the substantial part of
  the responses.}''} By comparing the proportions of liberal and
conservative responses in each of these clusters, we identify partisan
differences in how opinions are justified.

We cannot know for certain if the justifications produced by this method
truly reveal the causal antecedent of the initial responses generated --
this would require examining patterns of neural activation and
identifying how these patterns correspond to semantically coherent
features. In asking GPT-3 to justify its response, our approach reveals
the distribution of most likely justifications, reflecting both their
plausibility and acceptability. This is much like asking a human
respondent to justify a prior response. It does not necessarily provide
the true reason for the response, but it does source a network of
associated considerations that the respondent deems plausible to
themselves, acceptable to their imagined audiences, and relevant to the
issue at hand \parencite{Tourangeau2000-tk}. Just as these \emph{post hoc} justifications can still
provide insight into how an issue is understood by a human respondent,
it may similarly illuminate how the issue is ``understood'' by the
language model.

Collectively, our analytical approach is divided into three stages: (i)
validation of our method on well established political issues, (ii)
testing if GPT-3's representation of COVID-19 politicization anticipates
observed patterns, and (iii) exploration of the semantic considerations
underpinning GPT-3's predictions by asking simulated respondents to
explain their responses.

\subsubsection*{Validation}

To confirm that our prompt induces partisan differences as expected, we
conduct a series of validation tests on political issues already
well-established in 2019. Using the same style prompt described above
excluding only the passage about COVID-19, we generate liberal and
conservative responses to variously worded questions on topics of
abortion, climate change, gender and sexuality, race, immigration, drugs
and policing, gun control, healthcare, welfare state programs, and
business regulation. Across these 10 topical areas, we pose 37 distinct
questions, and each question had multiple wordings. For 35/37 questions,
the majority of wordings correctly predicted empirical partisan
differences on that issue. For one question, the association was in the
incorrect direction for 2/4 wordings, and for one question, no
association was identified. Results are presented in Appendix A. These
results also reveal systematic differences in the effect of question
wordings on stated positions. For example, asking a liberal- or
conservative-prompted model its ``stance'' on an issue almost always
brought the answer close to the center of the distribution, whereas when
an ``opinion'' is elicited, we find responses are more markedly
partisan.

These results provide confirmatory evidence that the prompts we designed
effectively induce partisan divides observable in American politics
circa 2019. This does not definitely establish that the model is equally
accurate in estimating American attitudes toward a hypothetical virus in
2019. The familiar political issues are within the training
distribution, whereas responding to questions about COVID-19 requires
out-of-distribution generalization. We have no ``ground truth'' for what
attitudes toward COVID-19 would have been in 2019. Indeed, if such a
data source existed we would not need to rely on simulation. But on
those attitudes for which validation is possible, we find encouraging
evidence that our prompts faithfully reproduce observable partisan
divisions.

\section*{RESULTS}

We begin by presenting results from prompts on vaccine-related topics.
Each of the figures below is a coefficient plot showing the effect of
``partisan priming'' on the classification scores for a given prompt's
outputs. Partisan priming is a binary variable coded 1 = ``liberal'', so
positive coefficients indicate that liberal prompts were more likely to
endorse the first anchoring term in the response dichotomy (e.g. ``good
idea''). This equivalently means that conservative prompts are more
likely than the liberal prompt to endorse the second anchoring term
(e.g. ``bad idea''). Coefficients with significant positive effects
(\emph{p}\textless0.05) are colored blue. Negative coefficients
conversely signify that conservatives are more likely to endorse the
first anchoring term than liberals. Coefficients with significant
negative effects are colored red.

\begin{figure}[!tp]
  \vspace{-0.5cm}
  \captionsetup{justification=raggedright,singlelinecheck=false}
  \caption{Coefficient plots from OLS models of partisanship predicting output classification scores on vaccine-related issues.}
  \centering
  \includegraphics[width=\textwidth, height=\textheight, keepaspectratio]{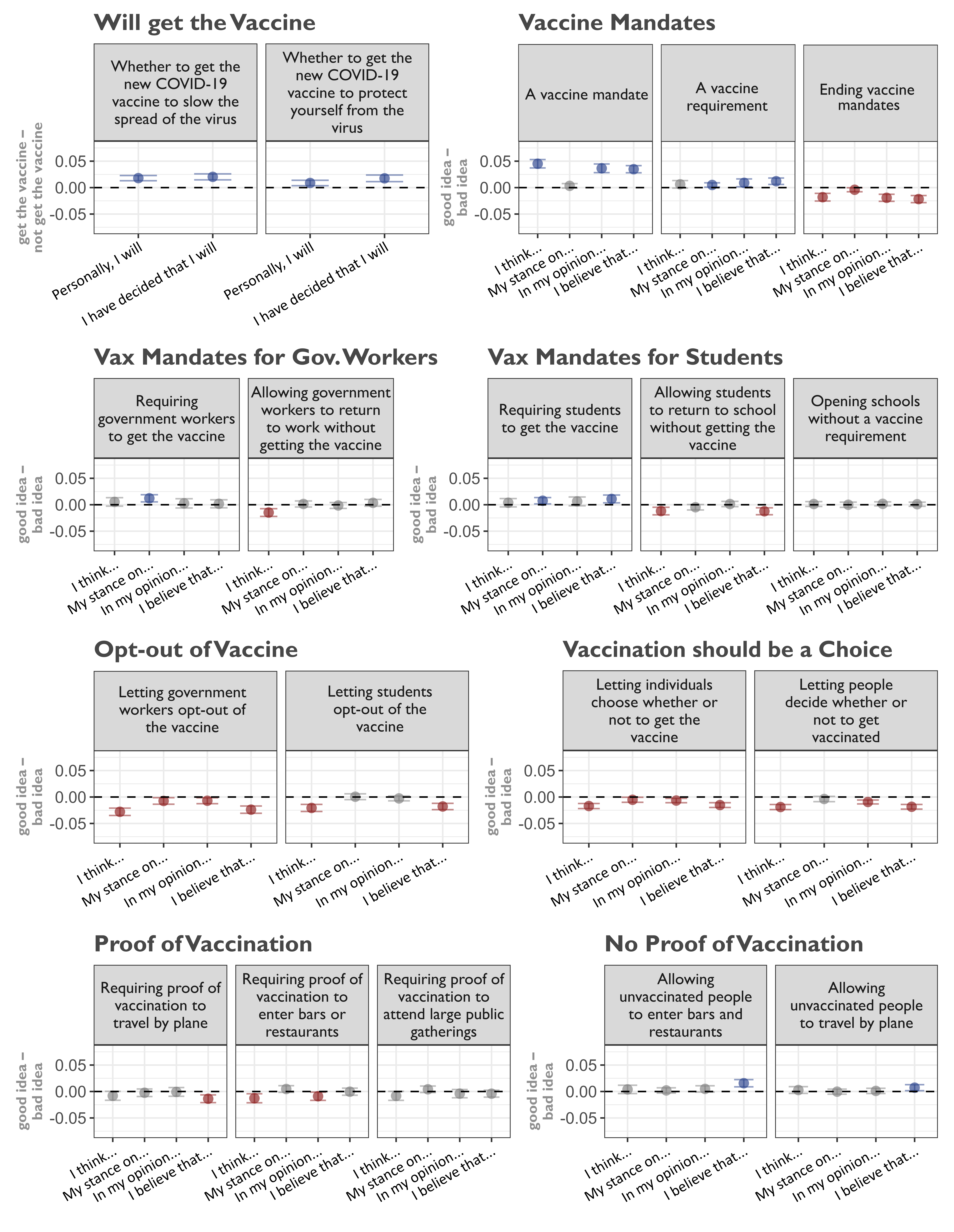}
  \label{fig:coefplot}
\end{figure}

\begin{figure}[!htbp]
  \captionsetup{justification=raggedright,singlelinecheck=false}
  \caption{Coefficient plots from OLS models of partisanship predicting output classification scores on mask related issues.}
  \label{fig:partisan2}
  \centering
  \includegraphics[width=0.7\textwidth]{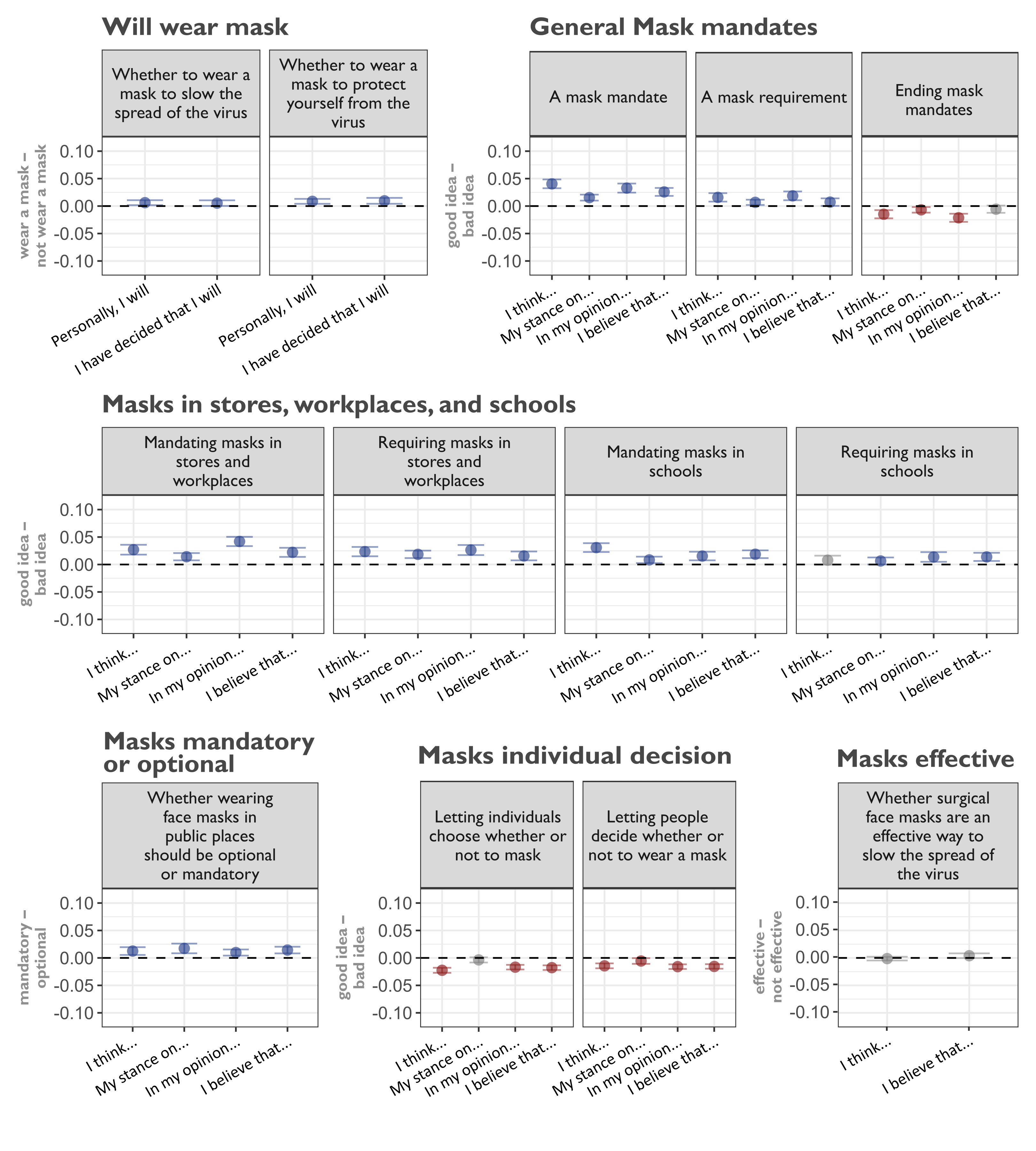}
\end{figure}

Figure \ref{fig:partisan2} displays coefficient plots of ``partisan priming'' predicting
output classification scores for vaccine-related issues. The plots in
the top left of Figure \ref{fig:coefplot} examine the effect of partisanship on the
intention to receive the vaccine. The positive coefficients in the first
plots show that liberal prompts were more likely to report an intention
to get the vaccine than conservative prompts. The effect is significant
for both wording variations (``Personally, I will'' and ``I have decided
that I will''), and for two question variations (``...to solve the
spread of the virus'' and ``...to protect yourself against the virus'').

The next plots show results to questions regarding vaccine mandates. In
the first question, liberal prompts are more likely to claim that a
vaccine mandate is a ``good idea'' than conservative prompts, with the
effect significant for three of four wordings. The second question
replaces the word ``mandate'' with ``requirement.'' The effect remains
positive for three of four wordings, although some effects are
attenuated. Conversely, when the question is changed to be about
\emph{ending} vaccine mandates, conservative prompts are more likely to
state that this is a ``good idea.''

In the second row of Figure \ref{fig:coefplot}, we test the effect of partisanship on
responses to vaccine mandates for government workers and students.
Although many of the coefficients fall short of statistical
significance, the significant effects are consistent with later observed
polarization: the liberal prompt is associated with stating that
``requiring government workers to get the vaccine'' is a good idea,
while conservative prompts is associated with saying ``allowing
government workers to return to work without getting the vaccine'' is a
good idea. Similarly, liberal prompts endorse ``requiring students to
get the vaccine'' while conservative prompts endorse ``allowing students
to return to school without getting the vaccine.'' However, when the
question is worded to probe a response to ``opening schools without a
vaccine requirement,'' no significant effect for partisanship is
identified.

The third row of Figure \ref{fig:coefplot} shows the effects of partisanship on responses
to whether getting the vaccine should be a choice. On the questions of
allowing government workers or students to ``opt-out'' of the vaccine,
conservative prompts were more likely in both cases to say this is a
good idea than liberal prompts. Similarly, the conservative prompts were
more likely than liberal prompts to say that ``letting individuals
choose'' and ``letting people decide'' to get the vaccine is a good
idea.

We see the first instances of model inferences being inconsistent with
historical observed patterns of polarization in the bottom row of Figure
5, where we ask about requiring proof of vaccination for various
activities. Conservative prompts are more likely to say it is a good
idea to require proof of vaccination to ``travel by plane'' or ``enter
bars or restaurants.'' The same pattern emerges when the question is
reworded, such that liberal prompts are more likely to state that
``allowing unvaccinated people'' to travel by plane or enter bars and
restaurants is a good idea.

We next run a similar set of tests for opinions regarding face mask
requirements. Results are presented in Figure \ref{fig:partisan2}. First, we see that
liberal prompts are more likely to express intent to wear face masks
both to slow the spread of the virus and for personal protection.
Liberal prompts were also more likely than conservative prompts to say
that mask ``mandates'' and mask ``requirements'' are good ideas.
Consistent with this, conservative prompts were more likely to respond
that ending mask mandates is a good idea.

The second row shows responses to prompts about requiring/mandating
masks in stores, workplaces, or schools. For both ``requiring'' and
``mandating'' wordings, and for both stores/workplaces and schools,
liberal prompts exhibit a greater likelihood of framing these measures
as a good idea.

In the third and final rows, we see results from prompts on whether
masking should be a personal choice and whether masks are effective. The
first plots show that conservative prompts were more likely to state
that ``letting individuals choose'' or ``letting people decide'' whether
to wear a mask is a good idea, with no apparent difference between these
wordings. The next plot shows that, when asked whether wearing masks in
public should be mandatory or optional, liberal prompts were more likely
to state that they should be ``mandatory'' than conservative prompts.
Lastly, partisanship had no statistically significant effect on
responses to whether face masks are an effective way to slow the spread
of the virus.

\begin{figure}[!htbp]
  \captionsetup{justification=raggedright,singlelinecheck=false}
  \caption{Coefficient plots from OLS models of partisanship predicting output classification scores on lockdown policies.}
  \label{fig:partisan3}
  \centering
\includegraphics[width=0.85\textwidth]{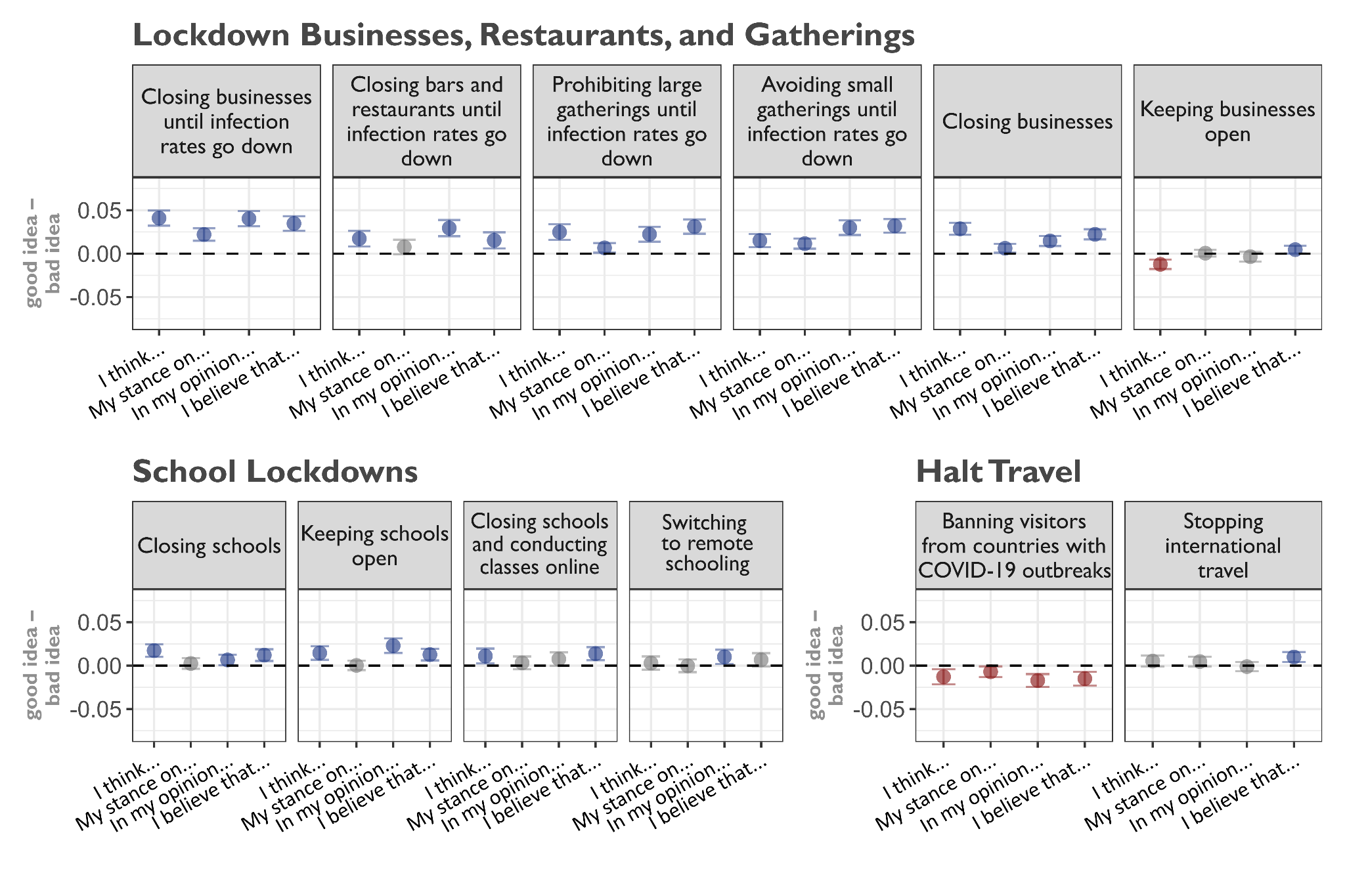}
\end{figure}

Next, in Figure \ref{fig:partisan3} we examine responses to questions about lockdowns. The
first two plots in the first row ask about closing business and closing
bars and restaurants. For both of these issues, the liberal prompts are
more likely to endorse the lockdowns. The next two issues concern
prohibiting large gatherings and avoiding small gatherings, and again
the liberal prompts are more likely to say each of these measures is a
``good idea.'' By contrast, the final question asks if keeping
businesses open is a good idea. The results for this question are mixed;
for completions beginning with ``I think,'' conservative prompts were
more likely to say keeping businesses open is a good idea. Yet for
completions beginning with ``I believe that,'' liberal prompts were more
likely to support the action. Although these results are mixed, they
still represent a movement in the correct direction relative to previous
questions.

In the second row of Figure \ref{fig:partisan3} are plots representing the effect of
partisanship on prompts regarding school lockdowns and halting
international travel. Results on the topic of school lockdowns are
mixed. Liberal prompts are more likely to say that ``keeping schools
open'' is a good idea than conservative prompts, but they are also more
likely to say ``closing schools and conducting classes online'' or
``switching to remote schooling'' are good ideas as well. Thus, we do
not see a clear partisan leaning in either direction in the question of
school lockdowns. Halting travel similarly exhibits some inconsistency.
Conservative prompts are more likely to endorse a ban on visitors from
countries with COVID-19 outbreaks, but are not more likely to endorse
``stopping international travel.'' Partisanship effects for ``stopping
international travel'' are non-significant, except liberal prompts were
more likely to endorse ``stopping international travel'' for completions
that began with ``I believe that.'' Results for halting travel are
therefore also inconsistent, but lean slightly in the correct
direction.\footnote{Gadarian et al. (2021) find that Democrats were less
  likely to support air travel restrictions or banning visitors from
  countries with COVID-19 outbreaks}

\begin{figure}[!htbp]
  \captionsetup{justification=raggedright,singlelinecheck=false}
  \caption{Coefficient plots from OLS models of partisanship predicting output classification scores on opinions toward the CDC and COVID-19 beliefs.}
  \label{fig:partisan4}
  \centering
  \includegraphics[width=0.65\textwidth]{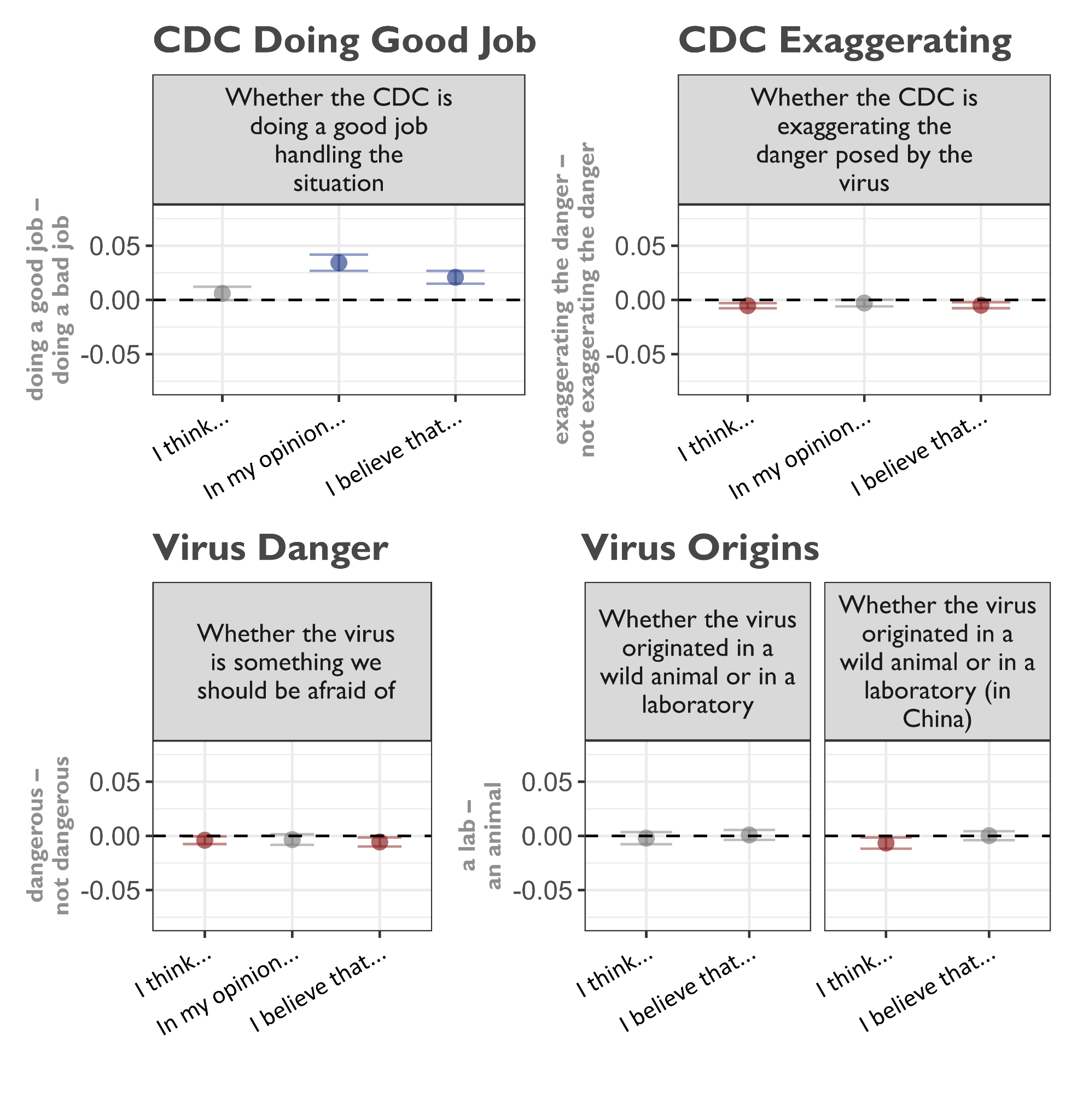}
\end{figure}

Lastly, Figure \ref{fig:partisan4} shows results from general questions on the COVID-19
pandemic. The first two questions assess confidence in the CDC. Liberal
prompts were more likely to state that the CDC is ``doing a good job''
handling the situation, while conservative prompts were more likely to
state that the CDC is ``exaggerating the danger posed by the virus.''
The next prompt poses whether the virus is something to be afraid of.
Intriguingly, conservative prompts were more likely to state that the
virus is dangerous than the liberal prompt, an apparent inconsistency
with the prior tendency of conservative prompts to state that the CDC is
exaggerating the threat. Finally, we ask whether the virus originated in
a lab or a wild animal. In these initial prompts, we do not observe a
partisan difference. However, when we mention in the prompt that
COVID-19 originated ``in China,'' conservative prompts are associated
with higher likelihood than liberal prompts of speculating that the
virus originated in a lab.

A collective overview of the results suggests that the partisan
associations generated by GPT-3 mirror historical associations at a rate
far outperforming chance. To assure that this assessment is correct, we
conduct multiple tests of overall statistical significance in the
Appendix. All tests, ranging from simple binomial tests to multi-level
cross-classified models all find that GPT-3's predictive capabilities
outperform chance with \emph{p}-values consistently below 0.001. Models
are described and results are presented in Appendix.

\subsubsection*{Generated Justifications}

GPT-3's success in anticipating the future politicization of COVID-19
suggests that the model encodes discursive associations from its
training texts that link liberalism with cautious responses to a novel
virus and conservatism with a rejection of such measures. However, the
preceding analyses tell us little about the nature of the associations
that facilitate the model's accurate predictions. To gain more detailed
insight into how GPT-3 links pandemic issues to political ideology, we
prompt the model to produce open-ended justifications for its previous
responses, then we cluster those justifications by theme. We divide
responses for each issue into two groups-- positive and negative -- and
use k-means clustering within each of these groups to identify clusters
of semantically similar justifications.

To provide the clearest insight into the system of relevant
considerations, we first analyze two cases in which the model predicted
particularly strong partisan divides: the intention to get vaccinated
and opinions toward mask mandates. We also apply this approach to one
issue where the model failed to correctly anticipate the direction of
COVID-19 polarization: whether the virus is something that should be
feared. While the model's incorrect predictions may simply be
``errors'', it is also possible that the model fails in situations where
the history of relevant considerations suggest a response that did not
manifest empirically. It is therefore plausible that incorrect
predictions indicate points where history deviated from what an analysis
of discourse alone would anticipate.

\begin{figure}[H]
  \captionsetup{justification=raggedright,singlelinecheck=false}
  \caption{Themes of justifications for getting / not getting
  the COVID-19 vaccine.}
  \label{fig:partisan5}
  \centering
  \includegraphics[width=\textwidth]{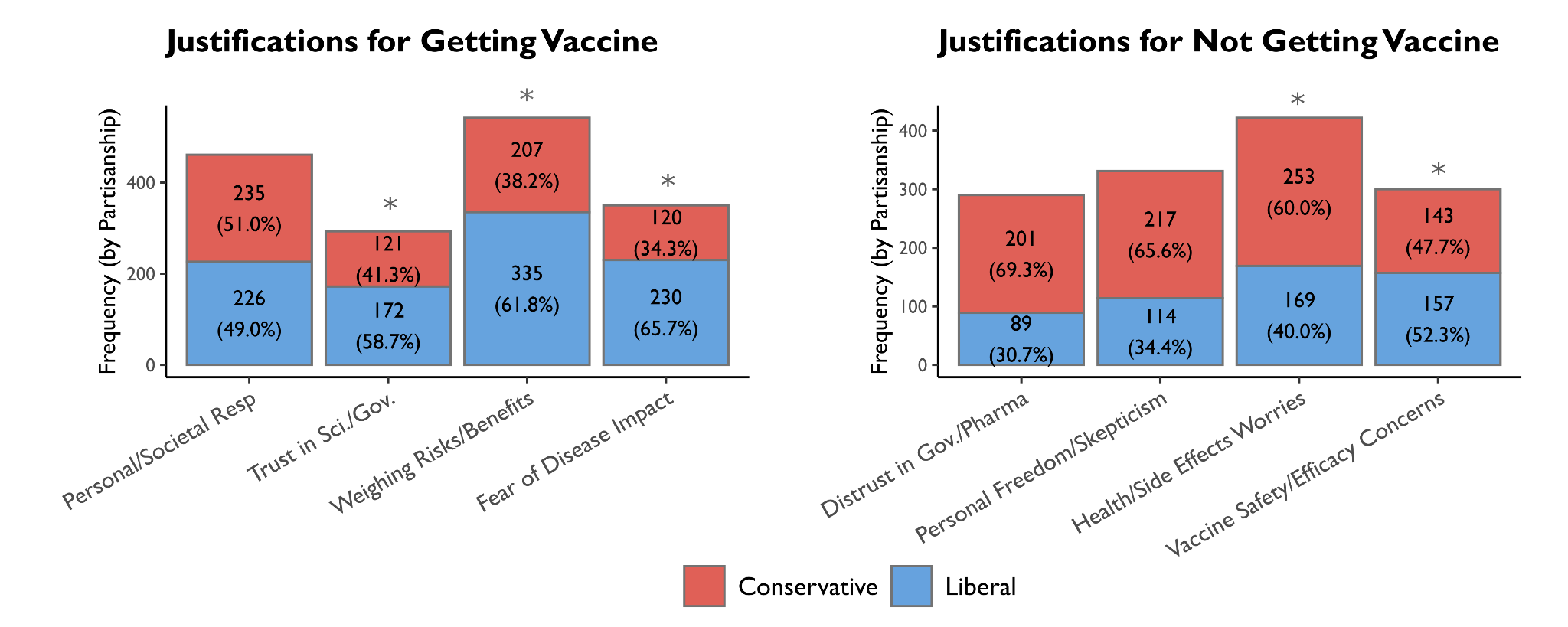}
  \smallskip 
  \small 
  \\ \textit{Note:} * p\textless0.05 two-sample binomial test that a cluster's composition is more liberal than the most conservative (left side) cluster. 
\end{figure}

Figure \ref{fig:partisan5} shows frequencies for each cluster of justifications in favor
and opposed to getting the vaccine. The first panel displays four
clusters of justifications for getting vaccinated, with the clusters in
each plot ordered from most conservative to most liberal. Although
liberal prompts disproportionately favor getting the vaccine,
conservative prompts comprise a slight majority in the first cluster of
justifications, Personal and Societal Responsibility (e.g. ``I think
that the American people have a duty to protect themselves and their
families.''). The remaining clusters are all majority liberal. In the
first of these clusters, Trust in Science and Government, the decision
to get vaccinated is justified by an appeal to the trustworthiness of
the scientists and government agencies involved in the vaccine's
production (e.g.``I have a lot of trust in the government to make sure
that the vaccine is safe and effective''). Justifications within the
Weighing Risks / Benefits cluster often argue that the vaccine carries
some risk but that this is far outweighed by the risk of contracting the
virus. The final category, Fear of Disease's Impact, emphasizes the
danger COVID-19 presents to both the respondent and the public at large.
In sum, an appeal to personal or societal responsibility is a
non-partisan justification for getting the vaccine, whereas appeals to
the trustworthiness of the institutions backing the vaccine as well as
appeals to safety tend to skew liberal.

The justifications for \emph{not} getting the vaccine also exhibit
substantial differences in their political skew. The most conservative
cluster is labeled Distrust in Government and Pharma (e.g. ``the CDC and
the FDA have been feeding us lies about the vaccine''). The following
cluster, Personal Freedom / Skepticism, expresses the belief that the
choice to vaccinate is highly personal (e.g ``I believe that the right
to choose what you put into your body is a fundamental human right'')
combined with occasional references to skepticism regarding the
vaccine's safety (e.g. ``I do not want to take the chance that it might
cause me harm, which is why I am against the government mandating that
everyone have this vaccine''). The final two clusters both highlight
potential side-effects of the vaccine (e.g. ``I believe the vaccine is
unsafe'') and concerns that the vaccine is either unsafe or ineffective
(e.g. ``the vaccine is still in its early stages and there are no
guarantees that it will work'' ) Interestingly, the Vaccine Safety /
Efficacy Concerns cluster is majority liberal, not conservative. This
suggests that the considerations that link conservatism to rejection of
the vaccine by GPT-3 are not simply ``anti-vax'' sentiments, given that
the partisan direction of this sentiment is equivocal.

\begin{figure}[!htbp]
  \captionsetup{justification=raggedright,singlelinecheck=false}
  \captionof{table}{Clusters of Justifications for Getting the Vaccine}
  \label{tab:table1}
  \centering
  \includegraphics[width=0.8\textwidth]{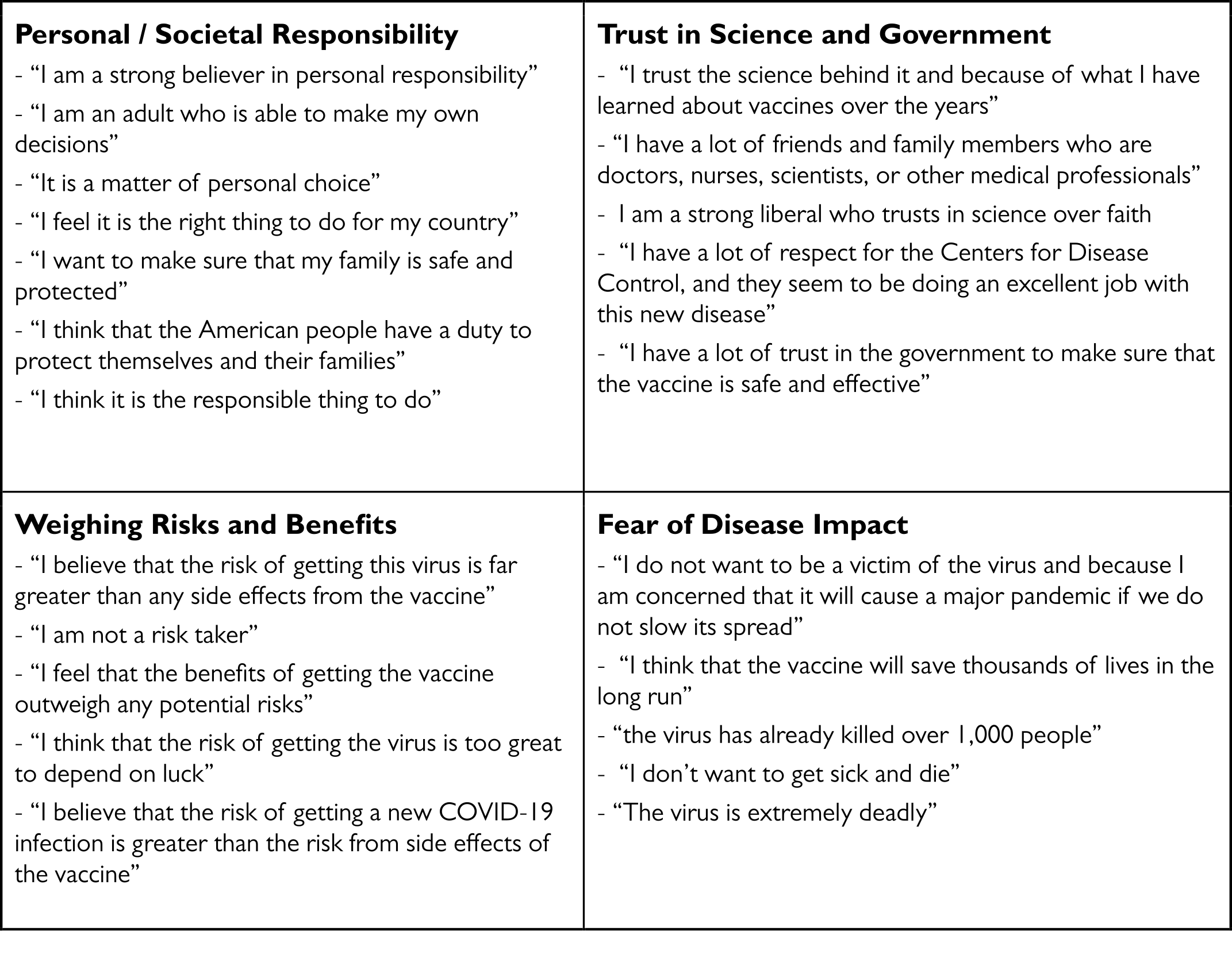}

  \vspace{1cm}

  \captionof{table}{Clusters for Justifications for Not Getting Vaccine}
  \label{tab:table2}
  \centering
  \includegraphics[width=0.8\textwidth]{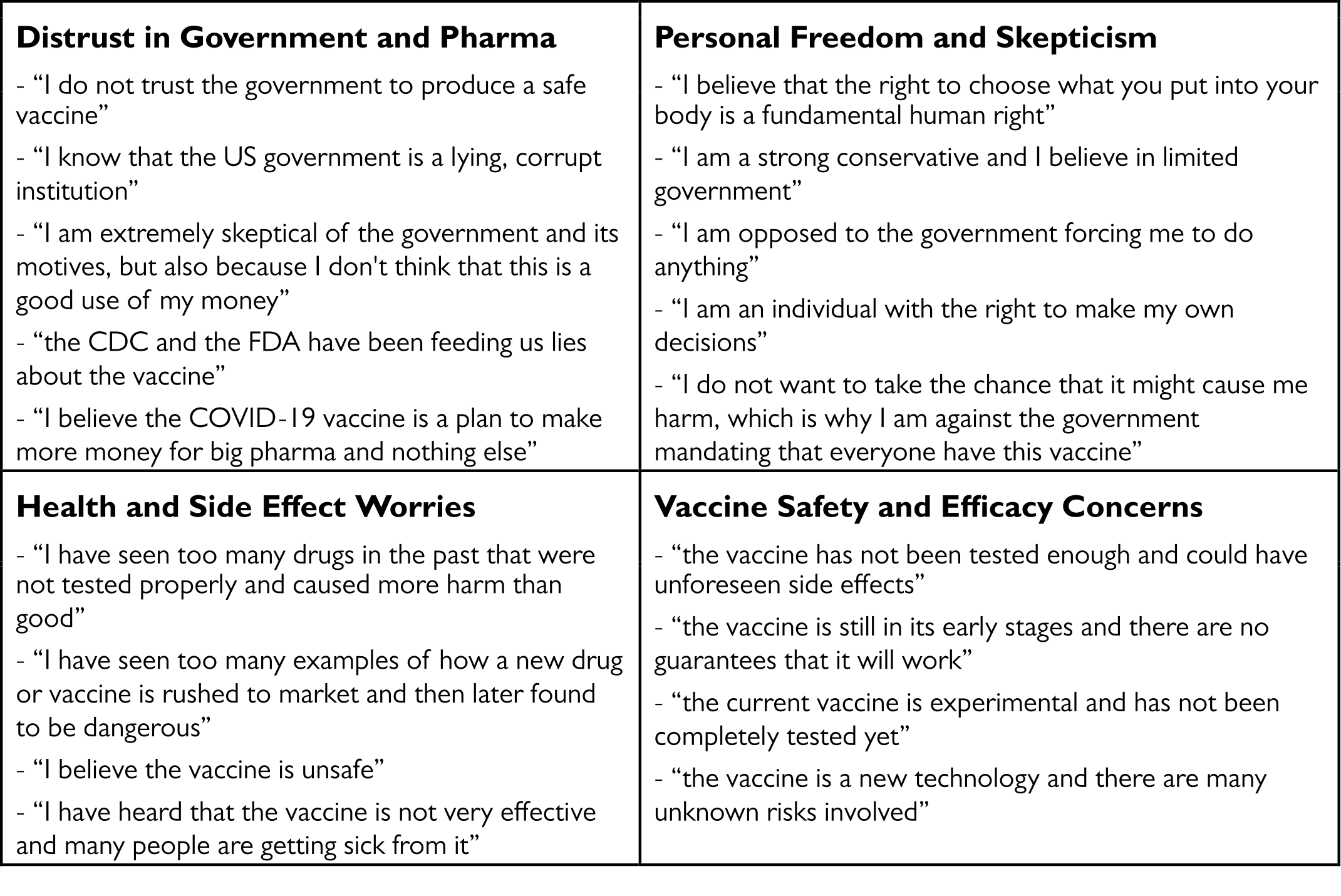}
\end{figure}

\begin{figure}[H]
  \captionsetup{justification=raggedright,singlelinecheck=false}
  \caption{Frequency of themes in justifications for and against Mask Mandates.}
  \label{fig:partisan6}
  \centering
  \includegraphics[width=\textwidth]{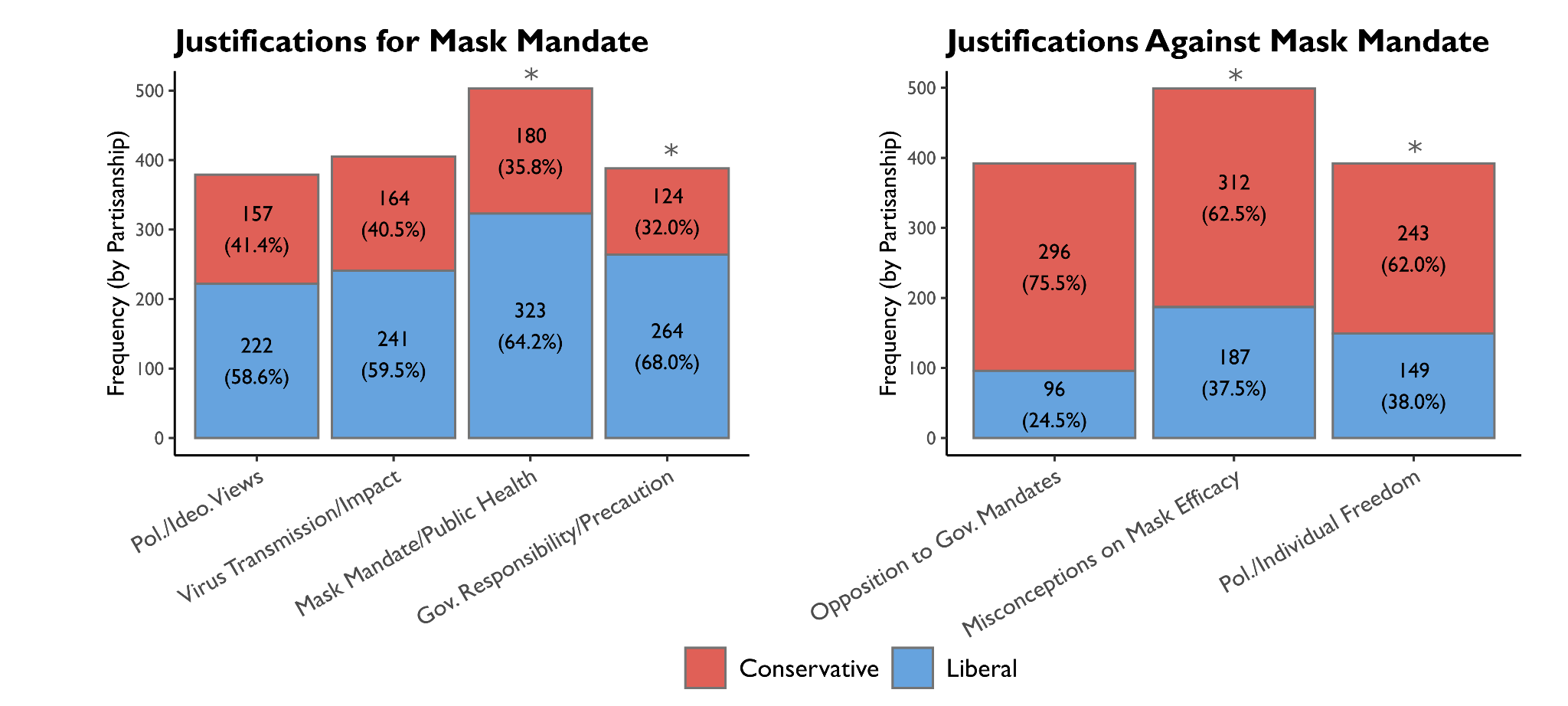}
  \smallskip 
  \small
  \\ \textit{Note:} * p\textless0.05 two-sample binomial test that a cluster's composition is more liberal than the most conservative (left side) cluster.
\end{figure}

Comparing these two plots, we see several themes that help to explain
partisan differences in intention to vaccinate. First, we observe a
juxtaposition between the liberal-skewed Trust in Science and Government
cluster in favor of vaccination and the highly conservative Distrust of
Government and Pharma cluster justifying vaccine hesitancy. References
to institutional trust among both positive and negative responses, along
with its corresponding partisan skew, suggests that GPT-3 identifies
confidence in government and scientific institutions as a major
consideration in the decision to vaccinate as well as a core issue
dividing liberals and conservatives. Another theme with partisan valence
is the appeal to safety vs. freedom. The most liberal justifications for
getting vaccinated emphasize dangers posed by the virus, and the most
liberal justification for \emph{not} getting vaccinated highlight
potential dangers of the vaccine. By contrast, the more conservative
themes emphasize personal responsibility and freedom, both of which
foreground choice over protection from harm.

Figure \ref{fig:partisan6} displays the clusters of justifications supporting and
opposing mask mandates. In the first panel, we see that liberal prompts
comprise the majority for all four clusters of justifications in favor
of mask mandates. However, some topics are more liberally skewed than
others. Explicit appeals to political ideology (e.g. ``I think the
Republicans have been wrong about this issue'') and comments about the
mode of transmission (e.g. ``when people sneeze or cough, they spread
the virus'') were only slightly tilted toward liberals. On the other
hand, appeals to the public health (e.g. ``a mask mandate would lower
the number of people who are exposed to this virus, which would protect
the public health and save lives'') and appeals to government
responsibility (e.g. ``the government has a responsibility to protect
the public'') are more disproportionately liberal.

Justifications \emph{against} mask mandates are divided into three
clusters, each of which is comprised of a majority of conservative
responses. The most conservative cluster expresses a general opposition
to government mandates (e.g. ``I believe that the government should not
be allowed to force people to wear masks''). The next cluster justifies
opposition to mask mandates by questioning masks' efficacy (e.g. ``a
face mask is not going to stop you from getting sick''). The final
cluster, which is still largely conservative, is characterized by
appeals to political and personal freedom (e.g. ``I believe that the
freedom to choose is an important right'').

\begin{figure}
  \captionsetup{justification=raggedright,singlelinecheck=false}
  \captionof{table}{Clusters of Justifications for Mask Mandate}
  \label{tab:masks}
  \centering
  \includegraphics[width=\textwidth]{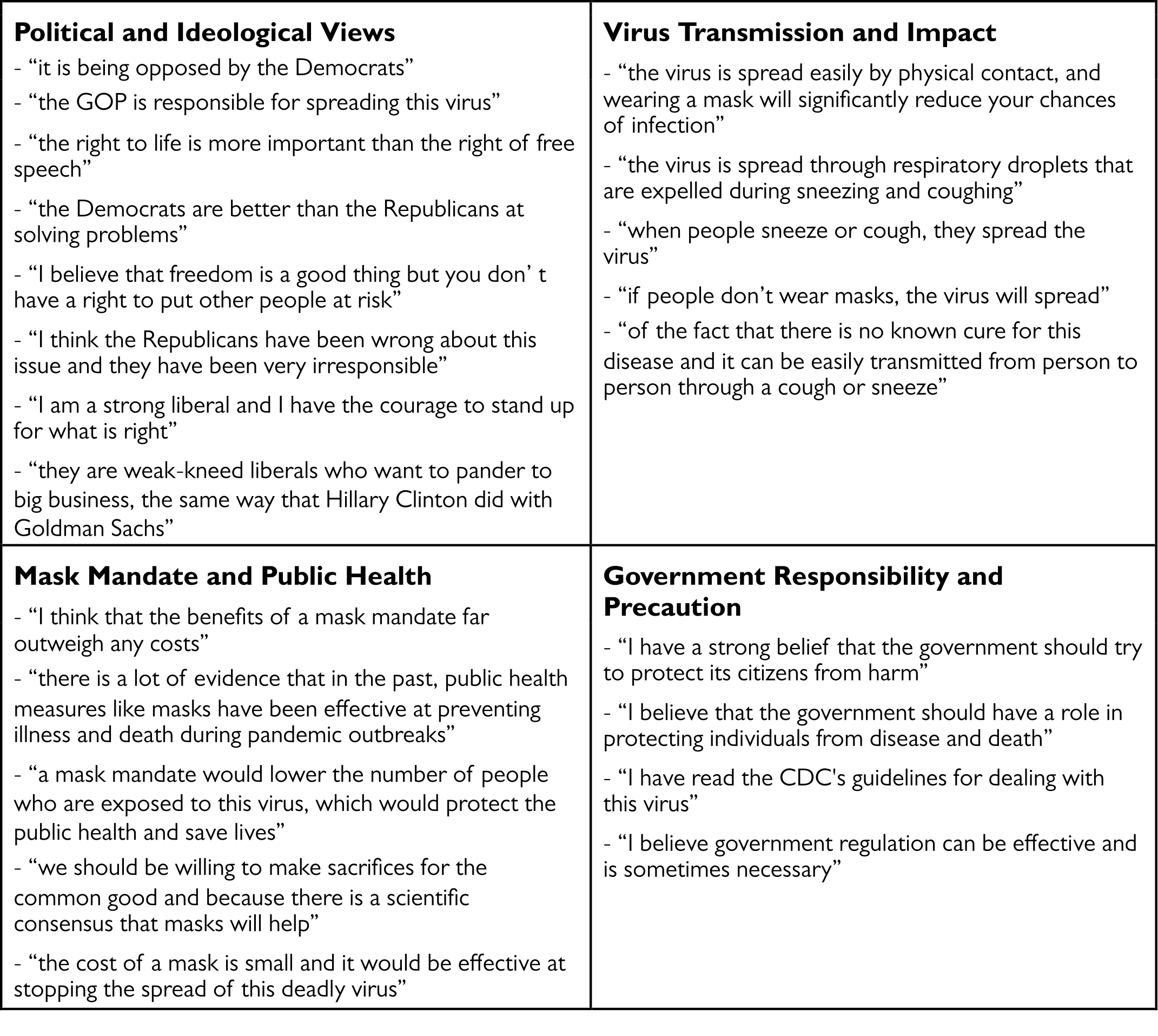}

  \vspace{1cm}
  
  \captionof{table}{Clusters for Justifications \textit{Against} Mask Mandate}
  \label{tab:nomasks}
  \centering
  \includegraphics[width=\textwidth]{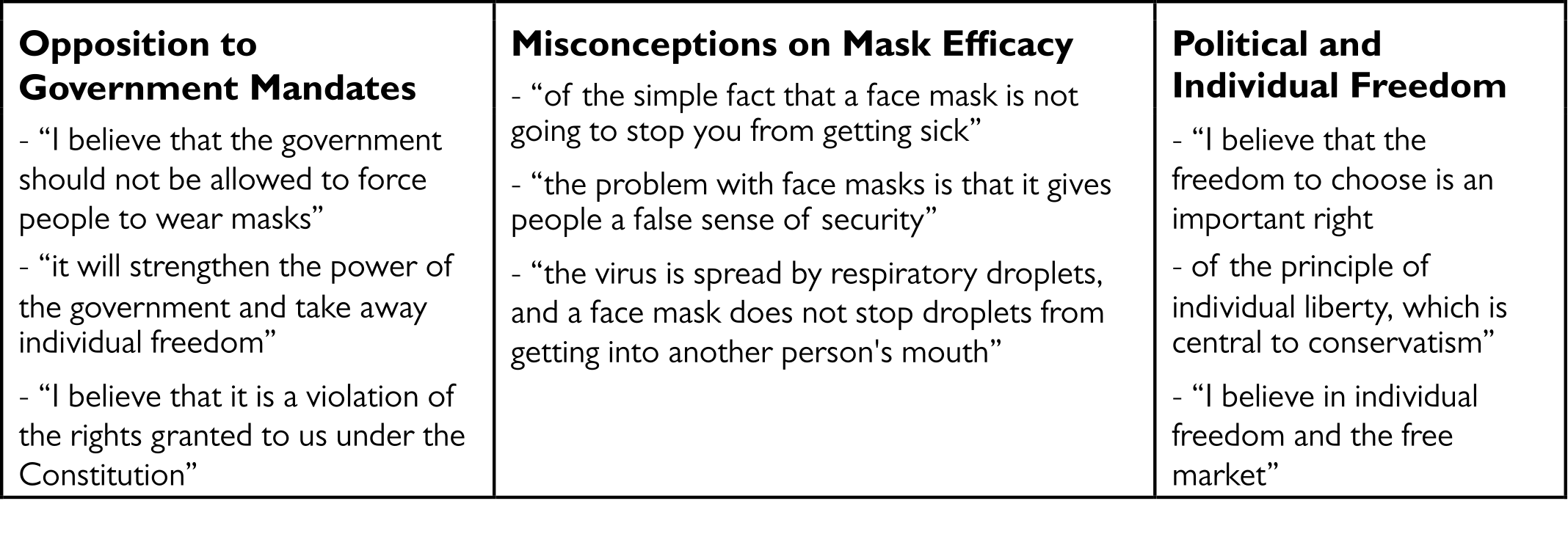}
\end{figure}

Taken together, these results further clarify the partisan division on mask mandates. One key
theme dividing liberal and conservative responses is their orientation toward government
intervention; the most liberal justification for mandates emphasizes the government's responsibility
to protect the public, whereas the most conservative justification for opposing mandates argues that
this measure would be government overreach. It is worth noting that this theme is closely related to
trust in the government, which previously emerged as a consideration dividing liberals and
conservatives in their intention to vaccinate. Similarly, we see the reappearance of partisan
differences in appeals to safety versus freedom. Liberal prompts are more likely to justify
their views by emphasizing the risk to public health, while conservative prompts disproportionately
highlight how such measures could jeopardize individual liberty. Lastly, we see a curious emergence
of partisan differences in the factual question of mask efficacy -- liberal responses argue that
masks capture respiratory droplets while conservative responses are skeptical of any protective effect.

\begin{figure}[!htbp]
  \captionsetup{justification=raggedright,singlelinecheck=false}
  \caption{Frequency of themes in justifications regarding fear COVID-19.}
  \label{fig:partisan7}
  \centering
  \includegraphics[width=\textwidth]{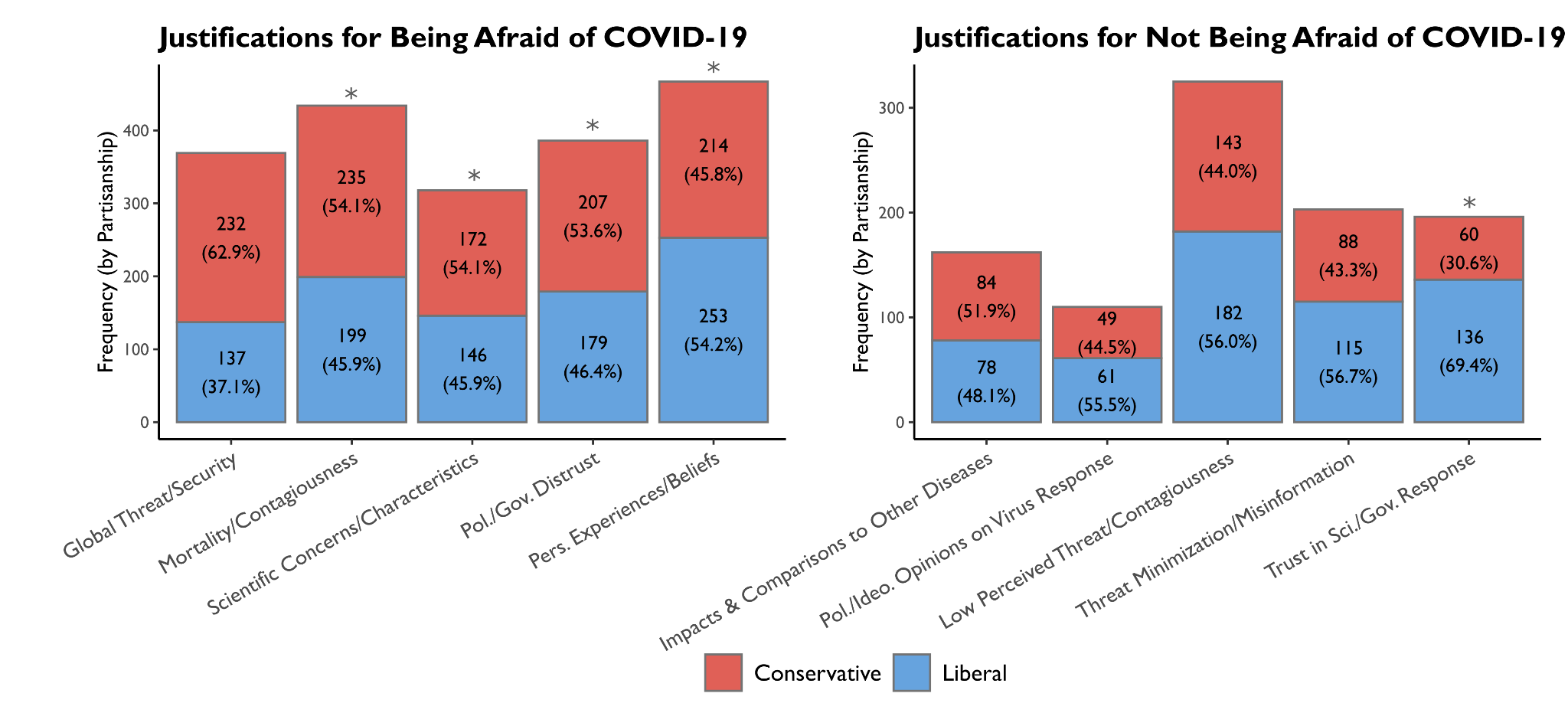}

  \vspace{1cm}

  \captionof{table}{Clusters for Justifications for Being Afraid of the COVID-19 Virus}
  \label{tab:afraid}
  \centering
  \includegraphics[width=\textwidth]{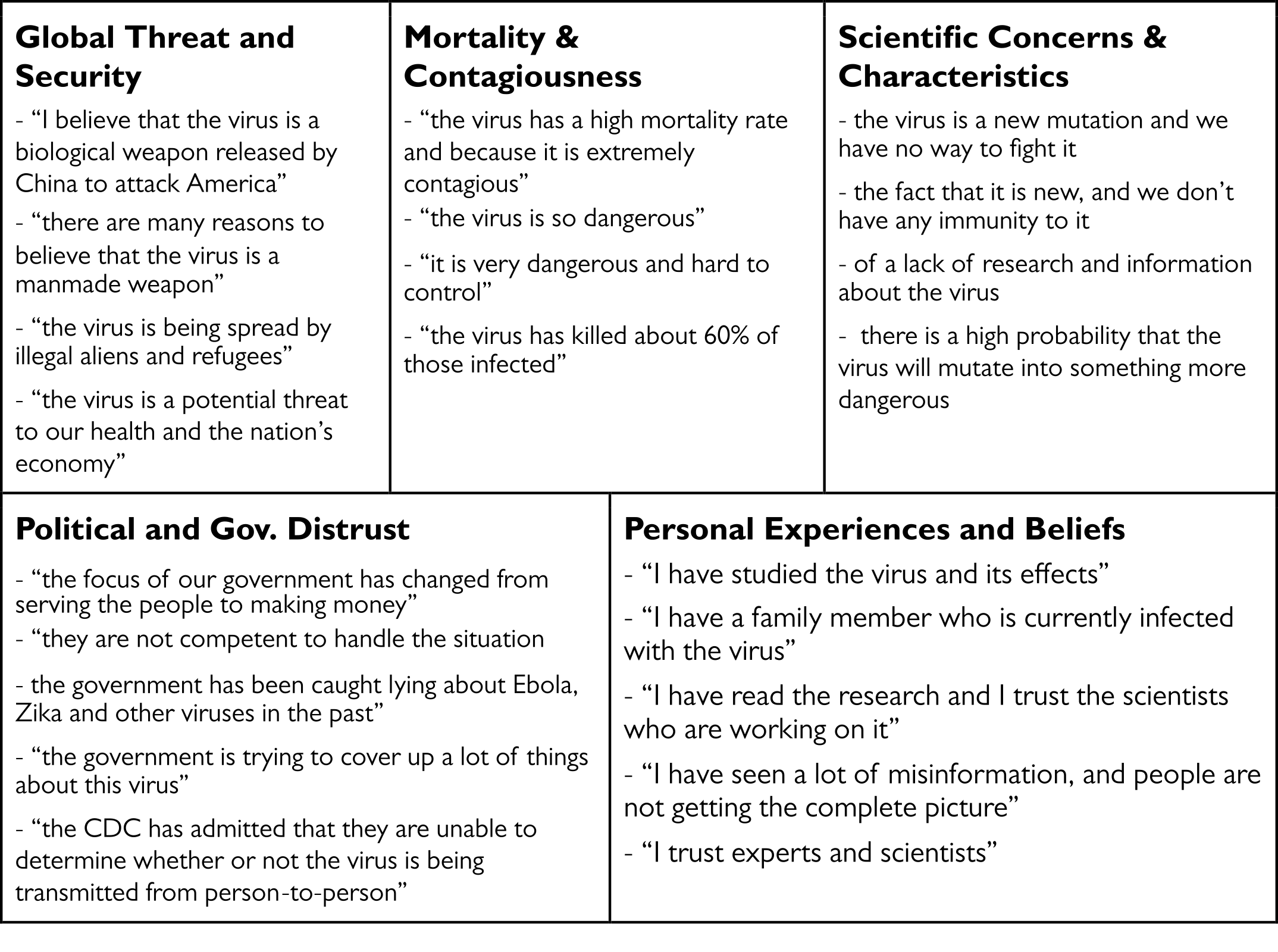}
\end{figure}

Finally, we pivot to examining an issue that was \emph{incorrectly}
forecast by the model -- expression of fear about the dangers of the
virus. The frequencies for each cluster of justifications are presented
in Figure \ref{fig:partisan7}. Justifications for fearing the virus are divided into five
clusters. The most conservative-skewed justifications frame the virus as
a threat to national security, commonly suggesting that it is a
bioweapon and linking it to foreign countries (e.g. ``I believe that the
virus is a biological weapon released by China to attack America'').
This is followed by concerns about the virus' lethality (e.g. ``the
virus has killed about 60\% of those infected''). The Scientific
Concerns and Characteristics cluster is also majority conservative and
emphasizes the virus's rapid mutation and science's inability to keep up
(e.g. ``{[}because{]} of a lack of research and information about the
virus''). The next two categories reveal a particularly instructive
partisan divide over whether the virus should be feared. Justifications
in the Political and Government Distrust category argue that the virus
should be feared because government institutions are untrustworthy and
incapable of handling the threat (e.g. ``they are not competent to
handle the situation''; ``the government is trying to cover up a lot of
things about this virus''). The most liberal cluster's LLM-generated
label (Personal Experiences \& Beliefs) is not particularly informative,
but from a sample of its justifications we see that it combines appeals
to personal experience (e.g. ``I have a family member who is currently
infected with the virus'') with appeals to scientific expertise (e.g.
``I trust experts and scientists'').

Lastly, the right panel of Figure \ref{fig:partisan7} displays clusters of justifications
for \emph{not} being afraid of the virus. The first cluster displays
near partisan parity, and argues that the new coronavirus is no worse
than other familiar viruses (e.g. ``the symptoms are very similar to the
flu''). The following three clusters which each display a slight liberal
skew view the virus as a political ploy (e.g. ``the pandemic is being
used to further conservative causes''), question the danger of the virus
(e.g. ``the virus is most likely to be harmless''), and argue that the
threat is overblown (e.g. ``the media has an interest in creating a
sense of panic''). The most intriguing results, however, come from the
most liberal-skewed cluster, Trust in Government and Science. Responses
in this cluster justify a less fearful response by expressing confidence
that the government and science will effectively protect the public from
the virus (e.g. ``I believe in the power of science to solve problems''
or ``I put my trust in the government and their scientists'').

We can thus see how a discursive association that led to accurate
forecasts for most issues may have supported incorrect associations in
this case; liberal's confidence in government and scientific
institutions could have dispelled fears about the virus by providing
assurance that the country is in good hands. Conversely, if COVID-19 had
been framed as a national security threat originating in a foreign
country, our results suggest that a cautious and vigilant response would
be consistent with contemporary conservative ideology. Of course, this
is not the path history took, but these alternative, imagined
justifications are consistent with the sophisticated and detailed model
of discourse encoded in GPT-3. To the extent that an LLM faithfully
preserves the principles for generating a historic discourse, even its
erroneous predictions may be illuminating.

\begin{table}
  \caption{Justifications for Not Being Afraid of the COVID-19 Virus}
  \label{tab:notafraid}
  \centering
  \includegraphics[width=\textwidth]{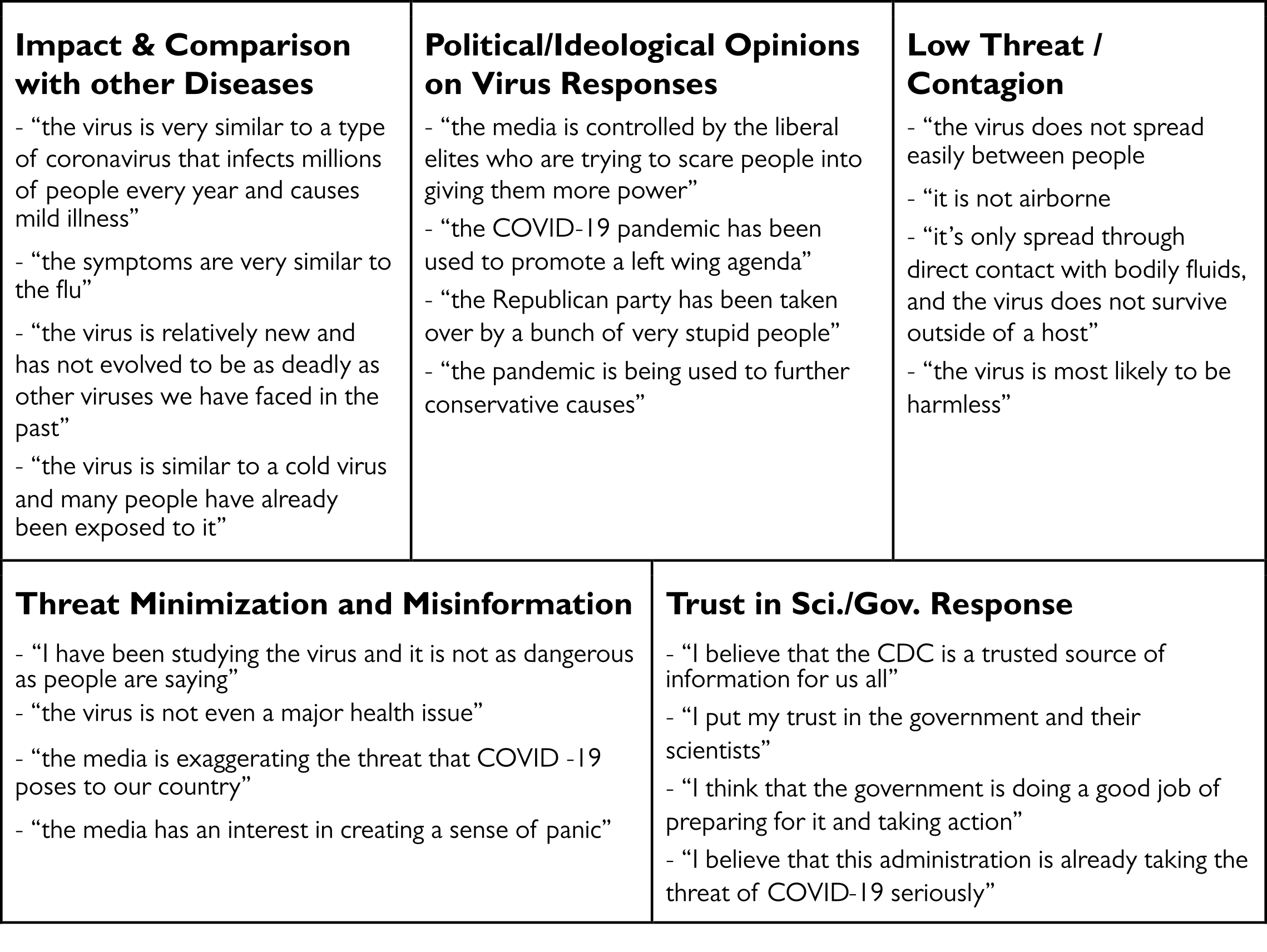}
\end{table}

\section*{Discussion}

LLMs can serve as powerful tools for the analysis of language and
culture. These models are unique in that they do not merely provide a
map of associations, they generate new texts consistent with the
linguistic patterns and discursive styles on which they were trained. It
is therefore possible to use LLMs to create ``digital doubles'' of
actors fluent in discourses included in the model's training texts. In
this study, we reconstructed the political opinion landscape of a
pivotal period, the year preceding the spread of COVID-19, to determine
whether the following politicization of the pandemic was predictable
given the existing regime of discourse and politics. We find that the
model predicts the correct direction of politicization far better than
chance across a wide array of pandemic-related issues. To gain insight
into how the model made these predictions, we prompt the LLM to produce
justifications for its responses. We find on key issues that the
distrust of institutions and the prioritization of personal freedom
characteristic of American conservatism corresponded with their greater
rates of rejecting sweeping policies to restrict the spread of the
virus, whereas liberal prompts tend to justify strong collective
responses to the virus with appeals to government responsibility and
public safety. These results suggest that the way the pandemic
politicized was largely consistent with existing repertoires of American
liberalism and conservatism, and that the pandemic was not an occasion
of substantial political innovation or surprise.

These findings speak to a fundamental question in the study of culture
-- to what extent are historical developments constrained by culture?
When a new issue emerges, is it interpreted within an existing system of
understanding, or is it ``up for grabs,'' with political and cultural
entrepreneurs offering competing frames for interpretation? Our findings
suggest that, in the case of COVID-19's polarization, existing schemas
of political sense-making steered the issue's reception. While President
Trump's early statements downplaying the need for drastic responses may
have helped crystallize the observed pattern of polarization, our
evidence suggests that this stance was consistent with a general
ideological tendency that predated the virus's emergence and channeled a
zeitgeist already present among Trump's constituency. We do not deny
that the political response to the COVID-19 pandemic could have unfolded
differently, but our evidence suggests that in this case, history
manifested along the predictable path consistent with the features of
liberalism and conservatism in 2019.

It is noteworthy that the considerations the LLM uses to justify its
stances often deviate from those emphasized within political psychology.
Many of the pillars of conservatism offered by political psychologists
--- uncertainty avoidance, fear of threat, need for order and structure,
and respect for authority
\parencite{Jost2006-wm} --- would have
likely motivated opposite responses to many COVID-19 policies.
Nevertheless, these considerations were not well represented among
produced justifications. Instead, LLM-generated justifications more
often made appeals to personal freedom and skepticism toward
governmental and scientific authorities, which were consistent with
conservatives' observed resistance to large-scale restrictions.

Our study presents one potential application of LLMs to the analysis of
cultural systems, but a variety of alternative approaches are likely to
prove fruitful for future inquiry. First, our analytic approach benefits
from historical happenstance --- the first human-level LLM was trained
on texts published immediately prior to an unprecedented socio-political
event. In order for LLMs to be a general tool for the analysis of
socio-cultural change, researchers should intentionally train models
sequentially on ordered time periods, releasing model ``checkpoints''
with training restricted to different periods of training texts. This
pattern of sequential training mirrors ``curriculum learning'' in which
training examples are presented to a model in sequence from simple to
complex \parencite{Bengio2009-ig}, but we propose a method that instead proceeds chronologically,
encouraging the model to learn the present in terms of the past, as in
the historic unfolding of social life. This sequential training approach
enables two key analytic opportunities. First, by comparing models
trained on subsequent periods, an analyst can identify the rate and
direction of historical cultural shifts. Second, by comparing periods
immediately prior and following moments of important social framing, the
influence of an event can be assessed.

\begin{figure}[!hp]
  \captionsetup{justification=raggedright,singlelinecheck=false}
  \caption{The causal opportunity associated with time-stamped LLM traces of cultural worlds and events.}
  \label{fig:future}
  \centering
  \includegraphics[width=\textwidth]{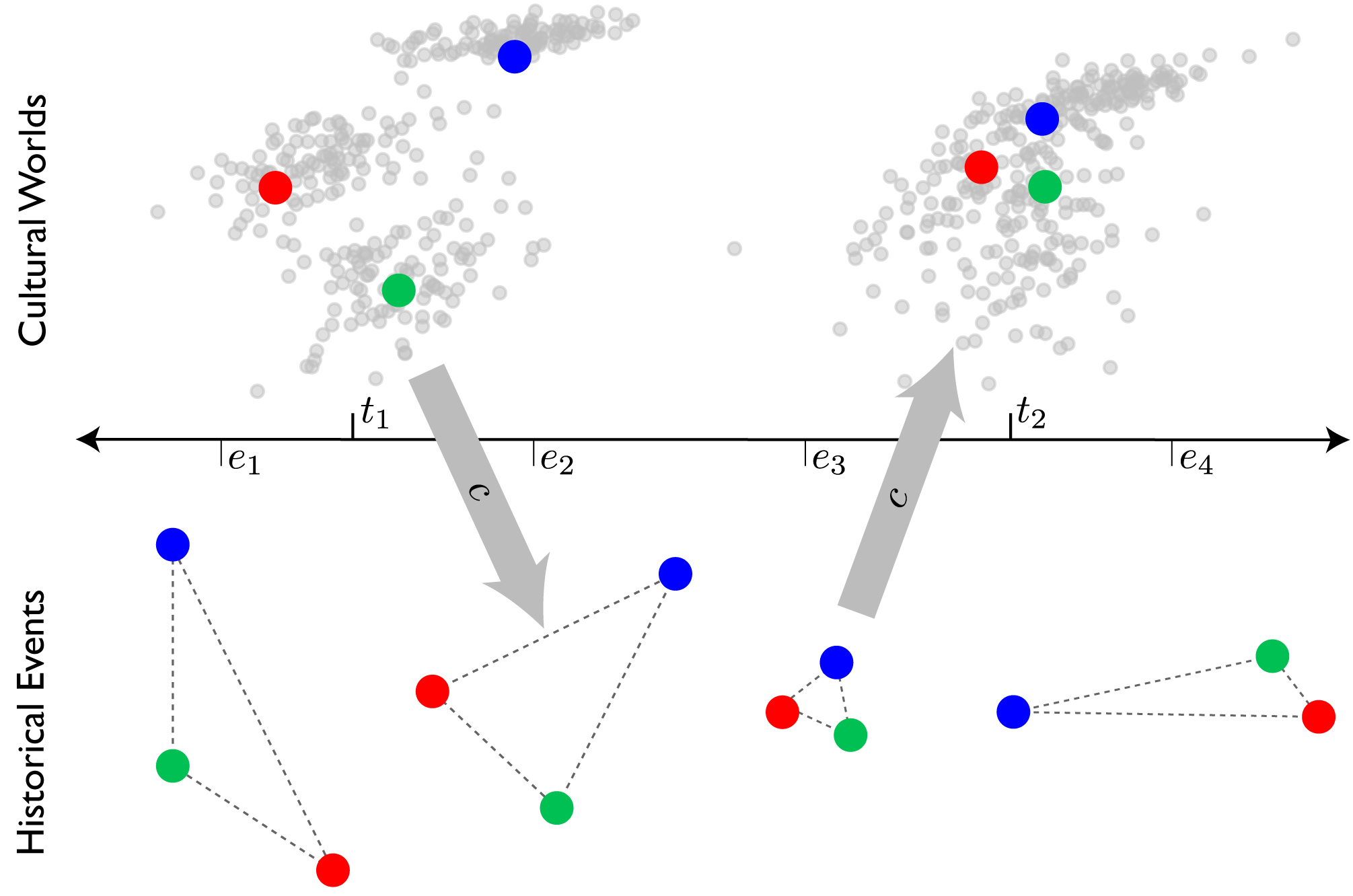}
\end{figure}

Figure \ref{fig:future} schematically represents how time-stamped, sequentially
trained LLMs could enable a ``causal cultural analysis.'' The top row
displays representations of cultural systems drawn from time-stamped
LLMs, here rendered as two dimensional point clouds. ``Historical
events'' are rendered below as networked associations between three
focal concepts (e.g., red, green, and blue). Linking these two levels of
observation could facilitate an empirical operationalization of Giddens'
structuration theory \parencite{Giddens1984-vx} or Coleman's ``boat'' models of macro-micro-macro relations
\parencite{Coleman1994-ay}. In our
example, cultural world models are checkpointed at two times:
\emph{t\textsubscript{1}} and \emph{t\textsubscript{2}}. Irregular
events \emph{e\textsubscript{1}} through \emph{e\textsubscript{4}} each
render the focal three concepts, which could represent ideas, facts,
stereotypes, or narrative elements with different orientations to one
another and the rest of the cultural world. In this scenario, event
\emph{e\textsubscript{2}} is wholly unsurprising, as it perfectly
reproduces the standing system of cultural relations. By contrast, event
\emph{e\textsubscript{3 }}not only deviates from the prior cultural
system, but changes the cultural world, becoming typical in the
subsequent time period. Most surprising associations (e.g.,
\emph{e\textsubscript{1}} and \emph{e\textsubscript{4}}) do not change
the cultural world. These represent the failed innovations, unfunny
jokes, and unfortunate accidents of history that do not cause anything
but their own forgetting.

Using LLMs as cultural world models, a wide range of causal
identification strategies
\parencite{Pearl2009-on} could be
deployed in order to do causal cultural history---to identify when
cultural events like speeches, concerts, new products, or viral memes
change the space of associations; and when the space of associations
changes the distribution of cultural events that follow. Our ability to
generate speech events from cultural world models, and evolve cultural
world models from speech events through fine-tuning enables the
production of richly situated counterfactuals for probabilistic
identification. This potential for cultural measurement and
identification could allow us to relax the assumed conservation of
influence from the macro-cultural ether to individual cultural behaviors
and back again, suggesting when change is conditioned and
disproportionately driven from the ``bottom-up'' or the ``top-down''.

We note that there are several practical reasons why LLMs are currently
not developed through chronological training, including: (1) the lower
average quality of earlier text, leading of path-dependent low-quality
models; (2) the inaccurately dated character of text on the web, LLM's
dominant corpus; (3) the phenomena of catastrophic forgetting whereby
deep neural networks learn new tasks without retaining the ability to
perform past ones
\parencite{McCloskey1989-rg, Ratcliff1990-bk}; and (4) the associated problem of mixing styles
that allows LLMs to efficiently link text across a large asynchronous
corpus. However, many of these limitations can be engineered around,
albeit with non-trivial expense. Early text can be corrected, or
translated into contemporary formats. Web-text can be dated using
time-stamped informal text and news. Earlier texts can be incorporated
into new training samples in proportion to metrics reflecting their
circulation and presence in current collective memory, such as their
recent citation rate.

Beyond this, LLMs present tremendous new opportunities to simulate
complex interactions between ``encultured'' agents. Our study uses LLMs
to generate simulated data that resemble responses to open-ended survey
questions, in which each response is treated as an independent sample
from a population of possible responses. But social life emerges through
interaction, and LLMs enable the simulation of semantically-rich
interaction at scale. Many forms of consequential social interactions
could be analyzed \emph{in silico} that would be challenging to
reproduce in a laboratory or online, such as parliamentary discourse,
scientific conferences, or cross-cultural collaborations. Moreover,
simulated interactions can be easily run millions of times over with
initial conditions experimentally manipulated to encourage wide coverage
of the space of possible interaction outcomes. Findings inductively
discovered from an extensive search over the interaction space could
then be validated with human subjects, or for certain hard-to-observe
social phenomena for which validation is impossible, simulation may
ultimately provide the best obtainable evidence.

Lastly, the internal representations of AI agents are directly
observable in a way that human internal representations are not, and may
therefore yield key insights into how complex meaning systems can be
efficiently modeled and compressed. Although the activation patterns of
deep neural networks are commonly described as black boxes, the growing
field of mechanistic interpretability is making progress in mapping
relations between neuronal activation and model behavior
\parencite{Bills2023-db, Elhage2022-bh}. More and more findings demonstrate the lower-level
``neurons'' with the transformer's self-attention architecture
facilitate novel interpretive and simulational capacities
\parencite{Hendel2023-ij}. As this field matures, it may enable
an ``AI neuroscience,'' whereby differences in behavior or discourse can
be translated into differences in latent representation. Although it is
likely that lessons from artificial neural networks cannot be directly
applied to human brains, their visibility and manipulability make them a
valuable site for exploratory studies linking distributed
representations to phenomena of culture and cognition.

As AI models continue to achieve more accurate approximations of human
behavior, they will become increasingly powerful tools for the analysis
of complex social processes. The advent of LLMs has now made simulation
studies of discourse possible for the first time, opening many new
avenues for investigation into language, culture, and meaning. Moreover,
just as LLMs trained on massive collections of text learn the patterns
of language, multi-modal models trained on images, videos, and other
records of social life can similarly distill underlying patterns across
these diverse domains
\parencite{Guilbeault2024-wz, Ludwig2024-nl}. Social scientists are fundamentally
interested in what people do, and direct observation of human action
remains the essential cornerstone of our science. But as empirically
realistic agents can be achieved \emph{in silico}, social simulations
will offer opportunities to explore social phenomena beyond the
observable, shedding light on the underlying patterns and branching
pathways that structure social life and give rise to its complex and
varied forms.

\newpage
\normalsize
\printbibliography

\cslsetup{class = in-text}

\cslcitation{Tesler2010-ty@1}{(\cslcite{Tesler2010-ty}{Tesler 2010})}
\cslcitation{Kaplan2020-lo@1}{(\cslcite{Kaplan2020-lo}{Kaplan et al. 2020})}
\cslcitation{Brown2020-nh@1}{(\cslcite{Brown2020-nh}{Brown et al. 2020})}
\cslcitation{DellAcqua2023-fo,Mollick2023-wy@1}{(\cslcite{DellAcqua2023-fo}{Dell’Acqua et al. 2023}; \cslcite{Mollick2023-wy}{Mollick and Mollick 2023})}
\cslcitation{Turing1950-in@1}{(\cslcite{Turing1950-in}{Turing 1950})}
\cslcitation{Ouyang2022-xw@1}{(\cslcite{Ouyang2022-xw}{Ouyang et al. 2022})}
\cslcitation{Argyle2023-ii,Kim2023-ji,Park2022-py@1}{(\cslcite{Argyle2023-ii}{Argyle et al. 2023}; \cslcite{Kim2023-ji}{Kim and Lee 2023}; \cslcite{Park2022-py}{Park et al. 2022})}
\cslcitation{Garcia-Pardina2022-cx@1}{(\cslcite{Garcia-Pardina2022-cx}{Garcia-Pardina et al. 2022})}
\cslcitation{Converse1964-kv,DellaPosta2020-ta,Hunzaker2019-ds@1}{(\cslcite{Converse1964-kv}{Converse 1964}; \cslcite{DellaPosta2020-ta}{DellaPosta 2020}; \cslcite{Hunzaker2019-ds}{Hunzaker and Valentino 2019})}
\cslcitation{Gadarian2021-su@1}{(\cslcite{Gadarian2021-su}{Gadarian, Goodman, and Pepinsky 2021})}
\cslcitation{OpenAI2023-wo@1}{(\cslcite{OpenAI2023-wo}{OpenAI 2023b})}
\cslcitation{Kaplan2008-nm@1}{(\cslcite{Kaplan2008-nm}{Kaplan 2008})}
\cslcitation{Ji2023-ut@1}{(\cslcite{Ji2023-ut}{Ji et al. 2023})}
\cslcitation{Kim2023-ji,Kozlowski2019-vh@1}{(\cslcite{Kim2023-ji}{Kim and Lee 2023}; \cslcite{Kozlowski2019-vh}{Kozlowski, Taddy, and Evans 2019})}
\cslcitation{Bonikowski2022-th,Grimmer2022-fu,Hannan2022-bs,Le_Mens2023-wa,Vicinanza_undated-cj@1}{(\cslcite{Bonikowski2022-th}{Bonikowski, Luo, and Stuhler 2022}; \cslcite{Grimmer2022-fu}{Grimmer, Roberts, and Stewart 2022}; \cslcite{Hannan2022-bs}{Hannan 2022}; \cslcite{Le_Mens2023-wa}{Le Mens et al. 2023}; \cslcite{Vicinanza_undated-cj}{Vicinanza, Goldberg, and Srivastava n.d.})}
\cslcitation{Senior2020-ee@1}{(\cslcite{Senior2020-ee}{Senior et al. 2020})}
\cslcitation{Jimenez-Luna2020-tg@1}{(\cslcite{Jimenez-Luna2020-tg}{Jiménez-Luna, Grisoni, and Schneider 2020})}
\cslcitation{Wilkins2023-mx,Zhou2018-cw@1}{(\cslcite{Wilkins2023-mx}{Wilkins 2023}; \cslcite{Zhou2018-cw}{Zhou et al. 2018})}
\cslcitation{Degrave2022-zi@1}{(\cslcite{Degrave2022-zi}{Degrave et al. 2022})}
\cslcitation{Jiao2021-mb,Mosavi2020-hy@1}{(\cslcite{Jiao2021-mb}{Jiao et al. 2021}; \cslcite{Mosavi2020-hy}{Mosavi et al. 2020})}
\cslcitation{Brynjolfsson2023-yu,Sourati2023-aa@1}{(\cslcite{Brynjolfsson2023-yu}{Brynjolfsson 2023}; \cslcite{Sourati2023-aa}{Sourati and Evans 2023})}
\cslcitation{Lai2024-qg@1}{(\cslcite{Lai2024-qg}{Lai et al. 2024})}
\cslcitation{Gentzkow2019-vu,Grimmer2022-fu@1}{(\cslcite{Gentzkow2019-vu}{Gentzkow, Kelly, and Taddy 2019}; \cslcite{Grimmer2022-fu}{Grimmer, Roberts, and Stewart 2022})}
\cslcitation{Mikolov2013-va,Pennington2014-dz@1}{(\cslcite{Mikolov2013-va}{Mikolov et al. 2013}; \cslcite{Pennington2014-dz}{Pennington, Socher, and Manning 2014})}
\cslcitation{Boutyline2023-sr,Garg2018-jf,Kozlowski2019-vh,Stoltz2019-mx@1}{(\cslcite{Boutyline2023-sr}{Boutyline, Arseniev-Koehler, and Cornell 2023}; \cslcite{Garg2018-jf}{Garg et al. 2018}; \cslcite{Kozlowski2019-vh}{Kozlowski, Taddy, and Evans 2019}; \cslcite{Stoltz2019-mx}{Stoltz and Taylor 2019})}
\cslcitation{Brown2020-nh@2}{(\cslcite{Brown2020-nh}{Brown et al. 2020})}
\cslcitation{Vaswani2017-wi@1}{(\cslcite{Vaswani2017-wi}{Vaswani et al. 2017})}
\cslcitation{Kaplan2020-lo@2}{(\cslcite{Kaplan2020-lo}{Kaplan et al. 2020})}
\cslcitation{Devlin2018-um,Mikolov2013-va@1}{(\cslcite{Devlin2018-um}{Devlin et al. 2018}; \cslcite{Mikolov2013-va}{Mikolov et al. 2013})}
\cslcitation{Sutskever2023-sa@1}{Sutskever (\cslcite{Sutskever2023-sa}{2023})}
\cslcitation{Gagniuc2017-xm,Hayes2013-yb@1}{(\cslcite{Gagniuc2017-xm}{Gagniuc 2017}; \cslcite{Hayes2013-yb}{Hayes 2013})}
\cslcitation{Shannon1951-ct@1}{(\cslcite{Shannon1951-ct}{Shannon 1951})}
\cslcitation{Rosenfeld2000-rq@1}{(\cslcite{Rosenfeld2000-rq}{Rosenfeld 2000})}
\cslcitation{Xue2020-qb@1}{(\cslcite{Xue2020-qb}{Xue et al. 2020})}
\cslcitation{Vaswani2017-wi@2}{(\cslcite{Vaswani2017-wi}{Vaswani et al. 2017})}
\cslcitation{Ouyang2022-xw@2}{(\cslcite{Ouyang2022-xw}{Ouyang et al. 2022})}
\cslcitation{Vaswani2017-wi@3}{(\cslcite{Vaswani2017-wi}{Vaswani et al. 2017})}
\cslcitation{Mikolov2013-va,Pennington2014-dz@2}{(\cslcite{Mikolov2013-va}{Mikolov et al. 2013}; \cslcite{Pennington2014-dz}{Pennington, Socher, and Manning 2014})}
\cslcitation{Argyle2023-ii@1}{(\cslcite{Argyle2023-ii}{Argyle et al. 2023})}
\cslcitation{Argyle2023-ii@2}{(\cslcite{Argyle2023-ii}{Argyle et al. 2023})}
\cslcitation{Brown2020-nh@3}{(\cslcite{Brown2020-nh}{Brown et al. 2020})}
\cslcitation{OpenAI2023-kd@1}{(\cslcite{OpenAI2023-kd}{OpenAI 2023a})}
\cslcitation{Ouyang2022-xw@3}{(\cslcite{Ouyang2022-xw}{Ouyang et al. 2022})}
\cslcitation{Argyle2023-ii@3}{(\cslcite{Argyle2023-ii}{Argyle et al. 2023})}
\cslcitation{Von_Oswald2023-xc@1}{(\cslcite{Von_Oswald2023-xc}{Von Oswald et al. 2023})}
\cslcitation{Dai2023-bg@1}{(\cslcite{Dai2023-bg}{Dai et al. 2023})}
\cslcitation{Douglas1966-re,Levi-Strauss1966-fw@1}{(\cslcite{Douglas1966-re}{Douglas 1966}; \cslcite{Levi-Strauss1966-fw}{Lévi-Strauss 1966})}
\cslcitation{Mead1942-zm@1}{(\cslcite{Mead1942-zm}{Mead 1942})}
\cslcitation{Swidler2003-il@1}{(\cslcite{Swidler2003-il}{Swidler 2003})}
\cslcitation{DiMaggio1997-rx@1}{(\cslcite{DiMaggio1997-rx}{DiMaggio 1997})}
\cslcitation{Baldassarri2014-mi,Boutyline2017-vj,Kiley2020-hi,Rawlings2020-yb,Swidler2003-il@1}{(\cslcite{Baldassarri2014-mi}{Baldassarri and Goldberg 2014}; \cslcite{Boutyline2017-vj}{Boutyline and Vaisey 2017}; \cslcite{Kiley2020-hi}{Kiley and Vaisey 2020}; \cslcite{Rawlings2020-yb}{Rawlings 2020}; \cslcite{Swidler2003-il}{Swidler 2003})}
\cslcitation{Feldman1992-lf,Goren2016-rw,Haidt2012-fy,Lakoff2010-yy@1}{(\cslcite{Feldman1992-lf}{Feldman and Zaller 1992}; \cslcite{Goren2016-rw}{Goren et al. 2016}; \cslcite{Haidt2012-fy}{Haidt 2012}; \cslcite{Lakoff2010-yy}{Lakoff 2010})}
\cslcitation{Carmines1989-yi,Zaller1992-eq@1}{(\cslcite{Carmines1989-yi}{Carmines and Stimson 1989}; \cslcite{Zaller1992-eq}{Zaller 1992})}
\cslcitation{Green2004-eq@1}{(\cslcite{Green2004-eq}{Green, Palmquist, and Schickler 2004})}
\cslcitation{Noel2014-es,Page2010-yj,Zaller1992-eq@1}{(\cslcite{Noel2014-es}{Noel 2014}; \cslcite{Page2010-yj}{Page and Shapiro 2010}; \cslcite{Zaller1992-eq}{Zaller 1992})}
\cslcitation{Mills1940-xp@1}{(\cslcite{Mills1940-xp}{Mills 1940})}
\cslcitation{Goffman2021-xf@1}{(\cslcite{Goffman2021-xf}{Goffman 2021})}
\cslcitation{Deane_undated-zp@1}{(\cslcite{Deane_undated-zp}{Deane, Parker, and Gramlich n.d.})}
\cslcitation{Stokes2020-vw@1}{(\cslcite{Stokes2020-vw}{Stokes et al. 2020})}
\cslcitation{Gadarian2021-su@2}{(\cslcite{Gadarian2021-su}{Gadarian, Goodman, and Pepinsky 2021})}
\cslcitation{Albrecht2022-kj,Allcott2020-iy,Chen2022-eb@1}{(\cslcite{Albrecht2022-kj}{Albrecht 2022}; \cslcite{Allcott2020-iy}{Allcott et al. 2020}; \cslcite{Chen2022-eb}{Chen and Karim 2022})}
\cslcitation{Homer2009-gw@1}{(\cslcite{Homer2009-gw}{Homer and French 2009})}
\cslcitation{Feldman1992-lf,Haidt2012-fy@1}{(\cslcite{Feldman1992-lf}{Feldman and Zaller 1992}; \cslcite{Haidt2012-fy}{Haidt 2012})}
\cslcitation{Jost2006-wm@1}{(\cslcite{Jost2006-wm}{Jost 2006})}
\cslcitation{Haidt2012-fy,Helzer2011-ni,Jost2017-sj,Oxley2008-pi,Terrizzi2013-po@1}{(\cslcite{Haidt2012-fy}{Haidt 2012}; \cslcite{Helzer2011-ni}{Helzer and Pizarro 2011}; \cslcite{Jost2017-sj}{Jost 2017}; \cslcite{Oxley2008-pi}{Oxley et al. 2008}; \cslcite{Terrizzi2013-po}{Terrizzi, Shook, and McDaniel 2013})}
\cslcitation{Pew_Research_Center2014-gi@1}{(\cslcite{Pew_Research_Center2014-gi}{Pew Research Center 2014})}
\cslcitation{Callaghan2019-zh,Colgrove2006-jy,Conis2014-qi,Jamison2019-yo@1}{(\cslcite{Callaghan2019-zh}{Callaghan et al. 2019}; \cslcite{Colgrove2006-jy}{Colgrove 2006}; \cslcite{Conis2014-qi}{Conis 2014}; \cslcite{Jamison2019-yo}{Jamison, Quinn, and Freimuth 2019})}
\cslcitation{Rosentiel2011-pi@1}{(\cslcite{Rosentiel2011-pi}{Rosentiel 2011})}
\cslcitation{DellaPosta2020-ta,Rawlings2022-pw@1}{(\cslcite{DellaPosta2020-ta}{DellaPosta 2020}; \cslcite{Rawlings2022-pw}{Rawlings 2022})}
\cslcitation{Furr2021-ud@1}{(\cslcite{Furr2021-ud}{Furr 2021})}
\cslcitation{Tourangeau2000-tk,Willis2004-lr@1}{(\cslcite{Tourangeau2000-tk}{Tourangeau, Rips, and Rasinski 2000}; \cslcite{Willis2004-lr}{Willis 2004})}
\cslcitation{Neelakantan2022-yd@1}{(\cslcite{Neelakantan2022-yd}{Neelakantan et al. 2022})}
\cslcitation{Neelakantan2022-yd@2}{(\cslcite{Neelakantan2022-yd}{Neelakantan et al. 2022})}
\cslcitation{Gadarian2021-su@3}{Gadarian, Goodman, and Pepinsky (\cslcite{Gadarian2021-su}{2021})}
\cslcitation{McCarthy2023-xj@1}{(\cslcite{McCarthy2023-xj}{McCarthy 2023})}
\cslcitation{Tourangeau2000-tk@1}{(\cslcite{Tourangeau2000-tk}{Tourangeau, Rips, and Rasinski 2000})}
\cslcitation{Jost2006-wm@2}{(\cslcite{Jost2006-wm}{Jost 2006})}
\cslcitation{Bengio2009-ig@1}{(\cslcite{Bengio2009-ig}{Bengio et al. 2009})}
\cslcitation{Giddens1984-vx@1}{(\cslcite{Giddens1984-vx}{Giddens 1984})}
\cslcitation{Coleman1994-ay@1}{(\cslcite{Coleman1994-ay}{Coleman 1994})}
\cslcitation{Pearl2009-on@1}{(\cslcite{Pearl2009-on}{Pearl 2009})}
\cslcitation{McCloskey1989-rg,Ratcliff1990-bk@1}{(\cslcite{McCloskey1989-rg}{McCloskey and Cohen 1989}; \cslcite{Ratcliff1990-bk}{Ratcliff 1990})}
\cslcitation{Bills2023-db,Elhage2022-bh@1}{(\cslcite{Bills2023-db}{Bills et al. 2023}; \cslcite{Elhage2022-bh}{Elhage et al. 2022})}
\cslcitation{Hendel2023-ij@1}{(\cslcite{Hendel2023-ij}{Hendel, Geva, and Globerson 2023})}
\cslcitation{Guilbeault2024-wz,Ludwig2024-nl@1}{(\cslcite{Guilbeault2024-wz}{Guilbeault et al. 2024}; \cslcite{Ludwig2024-nl}{Ludwig and Mullainathan 2024})}




\begin{thebibliography}{index = 1, hanging-indent = true, line-spacing = 1, entry-spacing = 1}

\bibitem{Albrecht2022-kj}
Albrecht, Don. 2022. “Vaccination, Politics and COVID-19 Impacts.” \textit{BMC Public Health} 22(1):96.

\bibitem{Allcott2020-iy}
Allcott, Hunt et al. 2020. “Polarization and Public Health: Partisan Differences in Social Distancing during the Coronavirus Pandemic.” \textit{Journal of Public Economics} 191:104254.

\bibitem{Argyle2023-ii}
Argyle, Lisa P. et al. 2023. “Out of One, Many: Using Language Models to Simulate Human Samples.” \textit{Political Analysis} 31:337–51.

\bibitem{Baldassarri2014-mi}
Baldassarri, Delia and Amir Goldberg. 2014. “Neither Ideologues nor Agnostics: Alternative Voters’ Belief System in an Age of Partisan Politics.” \textit{American Journal of Sociology} 120(1):45–95.

\bibitem{Bengio2009-ig}
Bengio, Yoshua, Jérôme Louradour, Ronan Collobert, and Jason Weston. 2009. “Curriculum Learning.” Pp. 41–48 in \textit{Proceedings of the 26th Annual International Conference on Machine Learning}, \textit{ICML ’09}. Montreal, Quebec, Canada: Association for Computing Machinery.

\bibitem{Bills2023-db}
Bills, Steven et al. 2023. “Language Models Can Explain Neurons in Language Models.” \textit{URL Https://Openaipublic. Blob. Core. Windows. Net/Neuron-Explainer/Paper/Index. Html. (Date Accessed: 14. 05. 2023)}.

\bibitem{Bonikowski2022-th}
Bonikowski, Bart, Yuchen Luo, and Oscar Stuhler. 2022. “Politics as Usual? Measuring Populism, Nationalism, and Authoritarianism in U.S. Presidential Campaigns (1952–2020) with Neural Language Models.” \textit{Sociological Methods \& Research} 51(4):1721–87.

\bibitem{Boutyline2023-sr}
Boutyline, Andrei, Alina Arseniev-Koehler, and Devin J. Cornell. 2023. “School, Studying, and Smarts: Gender Stereotypes and Education across 80 Years of American Print Media, 1930–2009.” \textit{Social Forces} 102(1):263–86.

\bibitem{Boutyline2017-vj}
Boutyline, Andrei and Stephen Vaisey. 2017. “Belief Network Analysis: A Relational Approach to Understanding the Structure of Attitudes.” \textit{The American Journal of Sociology} 122(5):1371–1447.

\bibitem{Brown2020-nh}
Brown, Tom B. et al. 2020. “Language Models Are Few-Shot Learners.” \textit{ArXiv} 2005.14165.

\bibitem{Brynjolfsson2023-yu}
Brynjolfsson, Erik. 2023. “The Turing Trap: The Promise \& Peril of Human-like Artificial Intelligence.” Pp. 103–16 in \textit{Augmented education in the global age}. Routledge.

\bibitem{Callaghan2019-zh}
Callaghan, Timothy, Matthew Motta, Steven Sylvester, Kristin Lunz Trujillo, and Christine Crudo Blackburn. 2019. “Parent Psychology and the Decision to Delay Childhood Vaccination.” \textit{Social Science \& Medicine (1982)} 238:112407.

\bibitem{Carmines1989-yi}
Carmines, Edward G. and James A. Stimson. 1989. \textit{Issue Evolution: Race and the Transformation of American Politics}. Princeton University Press.

\bibitem{Chen2022-eb}
Chen, Hsueh-Fen and Saleema A. Karim. 2022. “Relationship between Political Partisanship and COVID-19 Deaths: Future Implications for Public Health.” \textit{Journal of Public Health} 44(3):716–23.

\bibitem{Coleman1994-ay}
Coleman, James S. 1994. \textit{Foundations of Social Theory}. Harvard University Press.

\bibitem{Colgrove2006-jy}
Colgrove, James. 2006. \textit{State of Immunity: The Politics of Vaccination in Twentieth-Century America}. University of California Press.

\bibitem{Conis2014-qi}
Conis, Elena. 2014. \textit{Vaccine Nation}. University of Chicago Press.

\bibitem{Converse1964-kv}
Converse, Philip. 1964. “The Nature of Belief Systems in Mass Publics.” in \textit{Ideology and discontent}, edited by D. Apter. New York: Free Press.

\bibitem{Dai2023-bg}
Dai, Damai et al. 2023. “Why Can GPT Learn In-Context? Language Models Implicitly Perform Gradient Descent as Meta-Optimizers.”

\bibitem{Deane_undated-zp}
Deane, C., K. Parker, and J. Gramlich. n.d. “A Year of US Public Opinion on the Coronavirus Pandemic.”

\bibitem{Degrave2022-zi}
Degrave, Jonas et al. 2022. “Magnetic Control of Tokamak Plasmas through Deep Reinforcement Learning.” \textit{Nature} 602(7897):414–19.

\bibitem{DellAcqua2023-fo}
Dell’Acqua, Fabrizio et al. 2023. “Navigating the Jagged Technological Frontier: Field Experimental Evidence of the Effects of AI on Knowledge Worker Productivity and Quality.”

\bibitem{DellaPosta2020-ta}
DellaPosta, Daniel. 2020. “Pluralistic Collapse: The ‘Oil Spill’ Model of Mass Opinion Polarization.” \textit{American Sociological Review} 85(3):507–36.

\bibitem{Devlin2018-um}
Devlin, Jacob, Ming-Wei Chang, Kenton Lee, and Kristina Toutanova. 2018. “BERT: Pre-Training of Deep Bidirectional Transformers for Language Understanding.” \textit{ArXiv}.

\bibitem{DiMaggio1997-rx}
DiMaggio, Paul. 1997. “Culture and Cognition.” \textit{Annual Review of Sociology} 23(1):263–87.

\bibitem{Douglas1966-re}
Douglas, Mary. 1966. \textit{Purity and Danger: An Analysis of Concepts of Pollution and Taboo}. Routledge \& Kegan Paul.

\bibitem{Elhage2022-bh}
Elhage, Nelson et al. 2022. “Toy Models of Superposition.” \textit{ArXiv} 2209.10652.

\bibitem{Feldman1992-lf}
Feldman, Stanley and John Zaller. 1992. “The Political Culture of Ambivalence: Ideological Responses to the Welfare State.” \textit{American Journal of Political Science} 36(1):268–307.

\bibitem{Furr2021-ud}
Furr, Michael. 2021. \textit{Psychometrics: An Introduction}. SAGE Publications.

\bibitem{Gadarian2021-su}
Gadarian, Shana Kushner, Sara Wallace Goodman, and Thomas B. Pepinsky. 2021. “Partisanship, Health Behavior, and Policy Attitudes in the Early Stages of the COVID-19 Pandemic.” \textit{PloS One} 16(4):e0249596.

\bibitem{Gagniuc2017-xm}
Gagniuc, Paul A. 2017. \textit{Markov Chains: From Theory to Implementation and Experimentation}. John Wiley \& Sons.

\bibitem{Garcia-Pardina2022-cx}
Garcia-Pardina, Alejandro, Francisco José Abad, Alexander P. Christensen, Hudson Golino, and Luis Eduardo Garrido. 2022. “Dimensionality Assessment in the Presence of Wording Effects: A Network Psychometric and Factorial Approach.” \textit{PsyArXiv}.

\bibitem{Garg2018-jf}
Garg, Nikhil, Londa Schiebinger, Dan Jurafsky, and James Zou. 2018. “Word Embeddings Quantify 100 Years of Gender and Ethnic Stereotypes.” \textit{Proceedings of the National Academy of Sciences} 115(16):E3635–44.

\bibitem{Gentzkow2019-vu}
Gentzkow, Matthew, Bryan Kelly, and Matt Taddy. 2019. “Text as Data.” \textit{Journal of Economic Literature} 57(3):535–74.

\bibitem{Giddens1984-vx}
Giddens, Anthony. 1984. \textit{The Constitution of Society: Outline of the Theory of Structuration}. University of California Press.

\bibitem{Goffman2021-xf}
Goffman, Erving. 2021. \textit{The Presentation of Self in Everyday Life}. Knopf Doubleday Publishing Group.

\bibitem{Goren2016-rw}
Goren, Paul, Harald Schoen, Jason Reifler, Thomas Scotto, and William Chittick. 2016. “A Unified Theory of Value-Based Reasoning and U.S. Public Opinion.” \textit{Political Behavior} 38(4):977–97.

\bibitem{Green2004-eq}
Green, Donald P., Bradley Palmquist, and Eric Schickler. 2004. \textit{Partisan Hearts and Minds: Political Parties and the Social Identities of Voters}. Yale University Press.

\bibitem{Grimmer2022-fu}
Grimmer, Justin, Margaret E. Roberts, and Brandon M. Stewart. 2022. \textit{Text as Data: A New Framework for Machine Learning and the Social Sciences}. Princeton University Press.

\bibitem{Guilbeault2024-wz}
Guilbeault, Douglas et al. 2024. “Online Images Amplify Gender Bias.” \textit{Nature} 626(8001):1049–55.

\bibitem{Haidt2012-fy}
Haidt, Jonathan. 2012. \textit{The Righteous Mind: Why Good People Are Divided by Politics and Religion}. Knopf Doubleday Publishing Group.

\bibitem{Hannan2022-bs}
Hannan, Michael. 2022. “Measuring Memberships in Collectives in Light of Developments in Cognitive Science and Natural-Language Processing.” \textit{Sociological Science} 9:473–92.

\bibitem{Hayes2013-yb}
Hayes, Brian. 2013. “First Links in the Markov Chain.” \textit{American Scientist} 101(2):92.

\bibitem{Helzer2011-ni}
Helzer, Erik G. and David A. Pizarro. 2011. “Dirty Liberals! Reminders of Physical Cleanliness Influence Moral and Political Attitudes.” \textit{Psychological Science} 22(4):517–22.

\bibitem{Hendel2023-ij}
Hendel, Roee, Mor Geva, and Amir Globerson. 2023. “In-Context Learning Creates Task Vectors.” \textit{ArXiv} 2310.15916.

\bibitem{Homer2009-gw}
Homer, Jenny and Michael French. 2009. “Motorcycle Helmet Laws in the United States from 1990 to 2005: Politics and Public Health.” \textit{American Journal of Public Health} 99(3):415–23.

\bibitem{Hunzaker2019-ds}
Hunzaker, M. B. Fallin and Lauren Valentino. 2019. “Mapping Cultural Schemas: From Theory to Method.” \textit{American Sociological Review} 84(5):950–81.

\bibitem{Jamison2019-yo}
Jamison, Amelia M., Sandra Crouse Quinn, and Vicki S. Freimuth. 2019. “‘You Don’t Trust a Government Vaccine’: Narratives of Institutional Trust and Influenza Vaccination among African American and White Adults.” \textit{Social Science \& Medicine (1982)} 221:87–94.

\bibitem{Ji2023-ut}
Ji, Ziwei et al. 2023. “Towards Mitigating Hallucination in Large Language Models via Self-Reflection.” \textit{ArXiv} 2310.06271.

\bibitem{Jiao2021-mb}
Jiao, Yutao, Ping Wang, Dusit Niyato, Bin Lin, and Dong In Kim. 2021. “Toward an Automated Auction Framework for Wireless Federated Learning Services Market.” \textit{IEEE Transactions on Mobile Computing} 20(10):3034–48.

\bibitem{Jimenez-Luna2020-tg}
Jiménez-Luna, José, Francesca Grisoni, and Gisbert Schneider. 2020. “Drug Discovery with Explainable Artificial Intelligence.” \textit{Nature Machine Intelligence} 2(10):573–84.

\bibitem{Jost2006-wm}
Jost, John T. 2006. “The End of the End of Ideology.” \textit{The American Psychologist} 61(7):651–70.

\bibitem{Jost2017-sj}
Jost, John T. 2017. “Ideological Asymmetries and the Essence of Political Psychology.” \textit{Political Psychology} 38(2):167–208.

\bibitem{Kaplan2020-lo}
Kaplan, Jared et al. 2020. “Scaling Laws for Neural Language Models.” \textit{ArXiv} 2001.08361.

\bibitem{Kaplan2008-nm}
Kaplan, Sarah. 2008. “Framing Contests: Strategy Making under Uncertainty.” \textit{Organization Science} 19(5):729–52.

\bibitem{Kiley2020-hi}
Kiley, Kevin and Stephen Vaisey. 2020. “Measuring Stability and Change in Personal Culture Using Panel Data.” \textit{American Sociological Review} 85(3):477–506.

\bibitem{Kim2023-ji}
Kim, Junsol and Byungkyu Lee. 2023. “AI-Augmented Surveys: Leveraging Large Language Models for Opinion Prediction in Nationally Representative Surveys.” \textit{ArXiv} 2305.09620.

\bibitem{Kozlowski2019-vh}
Kozlowski, Austin C., Matt Taddy, and James A. Evans. 2019. “The Geometry of Culture: Analyzing the Meanings of Class through Word Embeddings.” \textit{American Sociological Review} 84(5):905–49.

\bibitem{Lai2024-qg}
Lai, Shiyang et al. 2024. “Evolving AI Collectives to Enhance Human Diversity and Enable Self-Regulation.” \textit{ArXiv} 2402.12590.

\bibitem{Lakoff2010-yy}
Lakoff, George. 2010. \textit{Moral Politics: How Liberals and Conservatives Think, Second Edition}. University of Chicago Press.

\bibitem{Le_Mens2023-wa}
Le Mens, Gaël, Balázs Kovács, Michael Hannan, and Guillem Pros. 2023. “Using Machine Learning to Uncover the Semantics of Concepts: How Well Do Typicality Measures Extracted from a BERT Text Classifier Match Human Judgments of Genre Typicality?” \textit{Sociological Science} 10:82–117.

\bibitem{Levi-Strauss1966-fw}
Lévi-Strauss, Claude. 1966. \textit{The Savage Mind}. University of Chicago Press.

\bibitem{Ludwig2024-nl}
Ludwig, J. and S. Mullainathan. 2024. “Machine Learning as a Tool for Hypothesis Generation.” \textit{The Quarterly Journal of Economics}.

\bibitem{McCarthy2023-xj}
McCarthy, Justin. 2023. “Roundup of Gallup COVID-19 Coverage.” \textit{Gallup.Com}. Retrieved (\url{https://web.archive.org/web/20230322222821/https://news.gallup.com/opinion/gallup/308126/roundup-gallup-covid-coverage.aspx}).

\bibitem{McCloskey1989-rg}
McCloskey, Michael and Neal J. Cohen. 1989. “Catastrophic Interference in Connectionist Networks: The Sequential Learning Problem.” Pp. 109–65 in \textit{Psychology of learning and motivation}, vol. 24, edited by G. H. Bower. Academic Press.

\bibitem{Mead1942-zm}
Mead, Margaret. 1942. \textit{And Keep Your Powder Dry}. New York: William Morrow.

\bibitem{Mikolov2013-va}
Mikolov, Tomas, Kai Chen, Greg Corrado, and Jeffrey Dean. 2013. “Efficient Estimation of Word Representations in Vector Space.” \textit{ArXiv} 1301.3781.

\bibitem{Mills1940-xp}
Mills, C. Wright. 1940. “Situated Actions and Vocabularies of Motive.” \textit{American Sociological Review} 5(6):904–13.

\bibitem{Mollick2023-wy}
Mollick, Ethan R. and Lilach Mollick. 2023. “Using AI to Implement Effective Teaching Strategies in Classrooms: Five Strategies, Including Prompts.”

\bibitem{Mosavi2020-hy}
Mosavi, Amir et al. 2020. \textit{Comprehensive Review of Deep Reinforcement Learning Methods and Applications in Economics}. SSRN.

\bibitem{Neelakantan2022-yd}
Neelakantan, Arvind et al. 2022. “Text and Code Embeddings by Contrastive Pre-Training.” \textit{ArXiv} 2201.10005.

\bibitem{Noel2014-es}
Noel, Hans. 2014. \textit{Political Ideologies and Political Parties in America}. Cambridge University Press.

\bibitem{OpenAI2023-kd}
OpenAI. 2023a. “GPT-4 Technical Report.” \textit{ArXiv} 2303.08774.

\bibitem{OpenAI2023-wo}
OpenAI. 2023b. “Models.” Retrieved (\url{https://platform.openai.com/docs/models/}).

\bibitem{Ouyang2022-xw}
Ouyang, Long et al. 2022. “Training Language Models to Follow Instructions with Human Feedback” edited by S. Koyejo et al. \textit{ArXiv} 2203.02155.

\bibitem{Oxley2008-pi}
Oxley, Douglas R. et al. 2008. “Political Attitudes Vary with Physiological Traits.” \textit{Science} 321(5896):1667–70.

\bibitem{Page2010-yj}
Page, Benjamin I. and Robert Y. Shapiro. 2010. \textit{The Rational Public: Fifty Years of Trends in Americans’ Policy Preferences}. University of Chicago Press.

\bibitem{Park2022-py}
Park, Joon Sung et al. 2022. “Social Simulacra: Creating Populated Prototypes for Social Computing Systems.” Pp. 1–18 in \textit{Proceedings of the 35th Annual ACM Symposium on User Interface Software and Technology}, \textit{UIST ’22}. Bend, OR, USA: Association for Computing Machinery.

\bibitem{Pearl2009-on}
Pearl, Judea. 2009. \textit{Causality}. Cambridge University Press.

\bibitem{Pennington2014-dz}
Pennington, J., R. Socher, and C. Manning. 2014. “Glove: Global Vectors for Word Representation.” \textit{Proceedings of the 2014}.

\bibitem{Pew_Research_Center2014-gi}
Pew Research Center. 2014. “Ebola Worries Rise, But Most Are ’fairly’ Confident in Government, Hospitals to Deal with Disease: Broad Support for US Efforts to Deal with Ebola in West Africa.”

\bibitem{Ratcliff1990-bk}
Ratcliff, R. 1990. “Connectionist Models of Recognition Memory: Constraints Imposed by Learning and Forgetting Functions.” \textit{Psychological Review} 97(2):285–308.

\bibitem{Rawlings2020-yb}
Rawlings, Craig M. 2020. “Cognitive Authority and the Constraint of Attitude Change in Groups.” \textit{American Sociological Review} 85(6):992–1021.

\bibitem{Rawlings2022-pw}
Rawlings, Craig M. 2022. “Becoming an Ideologue: Social Sorting and the Microfoundations of Polarization.” \textit{Sociological Science} 9:313–45.

\bibitem{Rosenfeld2000-rq}
Rosenfeld, R. 2000. “Two Decades of Statistical Language Modeling: Where Do We Go from Here?” \textit{Proceedings of the IEEE} 88(8):1270–78.

\bibitem{Rosentiel2011-pi}
Rosentiel, Tom. 2011. “Public Remains Divided over the Patriot Act.” Retrieved (\url{https://policycommons.net/artifacts/624569/public-remains-divided-over-the-patriot-act/1605857/}).

\bibitem{Senior2020-ee}
Senior, Andrew W. et al. 2020. “Improved Protein Structure Prediction Using Potentials from Deep Learning.” \textit{Nature} 577(7792):706–10.

\bibitem{Shannon1951-ct}
Shannon, C. E. 1951. “Prediction and Entropy of Printed English.” \textit{Bell System Technical Journal} 30(1):50–64.

\bibitem{Sourati2023-aa}
Sourati, Jamshid and James A. Evans. 2023. “Accelerating Science with Human-Aware Artificial Intelligence.” \textit{Nature Human Behaviour}.

\bibitem{Stokes2020-vw}
Stokes, Erin K. et al. 2020. “Coronavirus Disease 2019 Case Surveillance - United States, January 22-May 30, 2020.” \textit{MMWR. Morbidity and Mortality Weekly Report} 69(24):759–65.

\bibitem{Stoltz2019-mx}
Stoltz, Dustin S. and Marshall A. Taylor. 2019. “Concept Mover’s Distance: Measuring Concept Engagement via Word Embeddings in Texts.” \textit{Journal of Computational Social Science} 2(2):293–313.

\bibitem{Sutskever2023-sa}
Sutskever, Ilya. 2023. “An Observation on Generalization.”

\bibitem{Swidler2003-il}
Swidler, Ann. 2003. “Talk of Love: How Culture Matters.” \textit{Canadian Journal of Communication} 28(1).

\bibitem{Terrizzi2013-po}
Terrizzi, John A., Natalie J. Shook, and Michael A. McDaniel. 2013. “The Behavioral Immune System and Social Conservatism: A Meta-Analysis.” \textit{Evolution and Human Behavior : Official Journal of the Human Behavior and Evolution Society} 34(2):99–108.

\bibitem{Tesler2010-ty}
Tesler, Larry. 2010. “Adages and Coinages.” Retrieved (\url{https://web.archive.org/web/20100223000629/https://www.nomodes.com/Larry_Tesler_Consulting/Adages_and_Coinages.html}).

\bibitem{Tourangeau2000-tk}
Tourangeau, Roger, Lance J. Rips, and Kenneth Rasinski. 2000. \textit{The Psychology of Survey Response}. Cambridge University Press.

\bibitem{Turing1950-in}
Turing, A. M. 1950. “Computing Machinery and Intelligence.” \textit{Mind; a Quarterly Review of Psychology and Philosophy} LIX(236):433–60.

\bibitem{Vaswani2017-wi}
Vaswani, Ashish et al. 2017. “Attention Is All You Need.” \textit{Advances in Neural Information Processing Systems} 30.

\bibitem{Vicinanza_undated-cj}
Vicinanza, Paul, Amir Goldberg, and Sameer B. Srivastava. n.d. “Who Sees the Future? A Deep Learning Language Model Demonstrates the Vision Advantage of Being Small.”

\bibitem{Von_Oswald2023-xc}
Von Oswald, Johannes et al. 2023. “Transformers Learn In-Context by Gradient Descent.” Pp. 35151–74 in \textit{Proceedings of the 40th International Conference on Machine Learning}, vol. 202, \textit{Proceedings of Machine Learning Research}, edited by A. Krause et al. PMLR.

\bibitem{Wilkins2023-mx}
Wilkins, Alex. 2023. “AI Could Discover Wonder Materials.” \textit{New Scientist} 260(3467):8.

\bibitem{Willis2004-lr}
Willis, Gordon B. 2004. \textit{Cognitive Interviewing: A Tool for Improving Questionnaire Design}. SAGE Publications.

\bibitem{Xue2020-qb}
Xue, Linting et al. 2020. “mT5: A Massively Multilingual Pre-Trained Text-to-Text Transformer.” \textit{ArXiv} 2010.11934.

\bibitem{Zaller1992-eq}
Zaller, John. 1992. \textit{The Nature and Origins of Mass Opinion}. Cambridge University Press.

\bibitem{Zhou2018-cw}
Zhou, Quan et al. 2018. “Learning Atoms for Materials Discovery.” \textit{Proceedings of the National Academy of Sciences of the United States of America} 115(28):E6411–17.

\end{thebibliography}

\end{document}